\pdfoutput=1
\documentclass[11pt]{article}
\usepackage[margin=1in]{geometry}
\usepackage{amsmath,amsfonts,amssymb}
\usepackage{algorithmic}
\usepackage{textcomp}
\usepackage{url}
\usepackage{graphicx}
\usepackage{enumitem}
\usepackage{booktabs}
\usepackage{tabularx}
\usepackage{multirow}
\usepackage{array}
\usepackage{tikz}
\usetikzlibrary{arrows.meta,positioning}
\tikzset{
block/.style={draw,rounded corners,thick,align=center,minimum width=3.5cm,minimum height=1.2cm},
line/.style={-{Latex[length=3mm]},thick}
}
\usepackage{microtype}
\usepackage{float}
\usepackage{caption}
\usepackage[section]{placeins}
\usepackage[colorlinks=true,citecolor=blue,urlcolor=blue,linkcolor=blue]{hyperref}
\hypersetup{
pdftitle={When Linear RUL Labels Disagree with Vibration Degradation},
pdfauthor={Behrad Mousaei Shir-Mohammad; Seyed Reza Tavakoli; Siavash Ahmadi; Babak Hossein Khalaj; Mohammad Hossein Rohban; Mohammad Mohammadi},
pdfsubject={Stage-aware degradation-state target construction and bearing prognostics},
pdfkeywords={bearing prognostics, remaining useful life, health indicator, target design, vibration measurement, transformer, data fusion}
}
\newcommand{\affilnum}[1]{\textsuperscript{#1}}
\title{\textbf{When Linear RUL Labels Disagree with Vibration Degradation: A Stage-Aware Target and Dual-Scale Predictor Evaluated on XJTU-SY and IMS}}
\author{
\parbox{0.96\textwidth}{\centering
Behrad~Mousaei Shir-Mohammad\affilnum{1},\quad
Seyed~Reza~Tavakoli\affilnum{2}\\[0.25em]
Mohammad~Mohammadi\affilnum{2}\\[0.25em]
Siavash~Ahmadi\affilnum{5}\affilnum{*},\quad
Babak~Hossein~Khalaj\affilnum{3}\\[0.25em]
Mohammad~Hossein~Rohban\affilnum{4}\\[0.8em]
\footnotesize
\affilnum{1}School of Electrical and Computer Engineering, University of Tehran, Tehran, I.~R.~Iran.\\
\affilnum{2}Department of Mathematical Science, Sharif University of Technology, Tehran, I.~R.~Iran.\\
\affilnum{3}Department of Electrical Engineering, Sharif University of Technology, Tehran, I.~R.~Iran.\\
\affilnum{4}Department of Computer Engineering, Sharif University of Technology, Tehran, I.~R.~Iran.\\
\affilnum{5}Electronics Research Institute, Sharif University of Technology, Tehran, I.~R.~Iran.\\
\affilnum{*}Corresponding author: Siavash Ahmadi. Email: \href{mailto:s.ahmadi@sharif.edu}{s.ahmadi@sharif.edu}.\\
Mohammad Mohammadi: \href{mailto:mohammad.mohammadi97@sharif.edu}{mohammad.mohammadi97@sharif.edu}.
}
}
\date{}

\begin{document}
\maketitle

\begin{abstract}
Remaining useful life (RUL) studies commonly treat the label as fixed, although clock-linear labels may decline while measured vibration remains nearly stable and then changes rapidly near failure. We separate target design from prediction. A development-only pipeline constructs an oriented vibration health indicator, identifies chronological early--middle--late stages, and fits a continuous linear--quadratic--exponential degradation-state target. A compact CNN--LSTM and Transformer learn the target from causal feature sequences; validation-fitted Ordered Weighted Averaging combines their outputs. In a bearing-wise XJTU-SY hold-out, all bearings ending in ``5'' are excluded from fitted preprocessing, training, early stopping, and fusion. The fused predictor obtains RMSE $0.0608$, MAE $0.0392$, and $R^{2}=0.9617$, with the Transformer providing most of the accuracy. Target shape is assessed independently on three documented IMS failed-bearing trajectories. Against the best anchored linear fit to the same vibration-derived reference, the stage-aware curve reduces RMSE by $3.6$--$18.2\%$ and MAE by $3.1$--$31.1\%$; mean reductions are $10.2\%$ and $15.0\%$. Conservative BIC differences of $128.8$--$368.1$ favour the stage-aware representation, whereas moving-block bootstrap intervals cross zero. Thus, stage-dependent targets better describe the evaluated vibration-derived degradation states, but evidence remains descriptive with only three official IMS runs. The study establishes a measurement-oriented target-validity framework, not a universal nonlinear law for physical time-to-failure or robust cross-domain prediction.
\end{abstract}


\vspace{0.75em}
\hrule
\vspace{0.9em}

\section{Introduction}
\label{sec:introduction}
Remaining Useful Life (RUL) prediction is usually presented as a regression problem, but it is also a measurement and estimation problem because every result is conditional on the target supplied to the model. In bearing run-to-failure studies, that target is rarely observed directly. It is constructed from elapsed time, a failure threshold, a health indicator, or a combination of these quantities. Consequently, target construction is part of the prognostic model rather than a neutral data-preparation step. Reviews of machinery prognostics and health-indicator design similarly emphasise that useful predictions require both a defensible degradation representation and an evaluation protocol aligned with the intended maintenance decision \cite{gebraeel2023,zhou2022,li2024review}.

The most common benchmark label is the clock-linear curve
\begin{equation}
R_{\mathrm{clock}}(u)=1-u,\qquad u=t/T_f\in[0,1],
\end{equation}
where $T_f$ denotes the observed end of a run. This label is simple and interpretable as the fraction of test duration remaining. It is also linear by definition. The measured vibration condition, however, need not evolve linearly with clock time. Many bearing trajectories exhibit an extended weak-degradation interval, a transition regime, and rapid terminal deterioration. A clock-linear label can therefore require a model to predict degradation before the signal has changed and can obscure the terminal acceleration most relevant to maintenance. The scientifically defensible question is not whether physical time remaining, $T_f-t$, is nonlinear; it is whether the mapping from vibration condition to a normalised degradation-state target is better represented by one global line or by stage-dependent dynamics.

This distinction has become increasingly important as bearing RUL models grow more expressive. Health-indicator/GRU pipelines \cite{ni2023gru}, convolution-attention Transformers \cite{sun2024transformer}, local-enhancing Transformers \cite{peng2023let}, and parallel convolution--Transformer architectures \cite{tang2024parallel} can capture complex temporal patterns, but a powerful predictor cannot correct a conceptually unsuitable label. Piecewise and multistage degradation models explicitly recognise this issue \cite{qiu2023piecewise,wang2023twostage,guo2024wiener}. At the same time, survival models address censoring and output time-to-event distributions \cite{lillelund2025rulsurv}, transfer methods address operating-condition shift \cite{xu2025domain}, and probabilistic methods address predictive uncertainty \cite{bott2026uncertainty}. These approaches solve related but distinct problems; a target-shape study should not be presented as a direct replacement for them.

Recent papers in \emph{Measurement} further clarify the distinction between feature representation, state estimation, target definition, and transfer. Ayman et al. reviewed shallow and deep feature-learning methods for bearing prognostics, emphasising non-stationary vibration, class imbalance, and distribution shift \cite{ayman2025review}. Their taxonomy explains how temporal, spatial, and spatiotemporal representations affect downstream RUL prediction. The review nevertheless treats the evaluation target as given rather than testing whether it is consistent with the measured degradation trajectory.

Li et al. constructed a dynamically normalised health indicator and combined it with Bayesian recurrent state estimation for high-speed wind-turbine bearings \cite{li2025dynamic}. This directly connects vibration-condition representation with sequential state estimation and is close to the measurement perspective adopted here. Its main emphasis is estimation after health-indicator construction, whereas the present study separately tests the shape of the target fitted to that indicator.

Shuang et al. proposed a multisource--multitarget domain-adaptation network for bearing RUL prediction across equipment and operating conditions \cite{shuang2024msmt}. Their one-total/multi-branch design aims to retain domain-invariant and domain-specific information and improve cross-equipment generalisation. This addresses transfer of a predictor, but not whether the supervised RUL label itself agrees with the observed degradation state.

Zhang et al. introduced DCDAN, combining wavelet-based data enhancement, multiscale temporal convolution, attention-enhanced recurrent modelling, and domain-discrepancy reduction \cite{zhang2025dcdan}. The method explicitly represents degradation features at multiple scales and improves cross-domain prediction on two bearing datasets. It remains predictor- and transfer-centred, whereas the present work tests target adequacy before interpreting predictor accuracy.

The relative advantage of the proposed framework is therefore specific rather than universal. It is better suited to the earlier measurement-validity question because it compares stage-aware and anchored-linear representations against the same vibration-derived reference on a second rig, excludes final-test bearings from all fitted development steps, and reports BIC together with a dependence-aware bootstrap. This design reduces the risk that a strong predictor masks a mismatched label and makes the contribution complementary to recent \emph{Measurement} methods rather than a replacement for their transfer and state-estimation capabilities.

The present work is organised around three research questions:
\begin{enumerate}[leftmargin=*]
    \item \textbf{RQ1 -- Target adequacy:} Does a continuous three-stage target describe a vibration-derived degradation reference better than the strongest comparable global linear target on an independent bearing test rig?
    \item \textbf{RQ2 -- Predictive learnability:} Can causal feature sequences predict the proposed target on bearings excluded from all fitted development steps?
    \item \textbf{RQ3 -- Added value and limits:} How much is contributed by dual-scale model fusion, and does target choice alone resolve cross-run distribution shift?
\end{enumerate}

\begin{figure*}[!t]
    \centering
    \includegraphics[width=\textwidth]{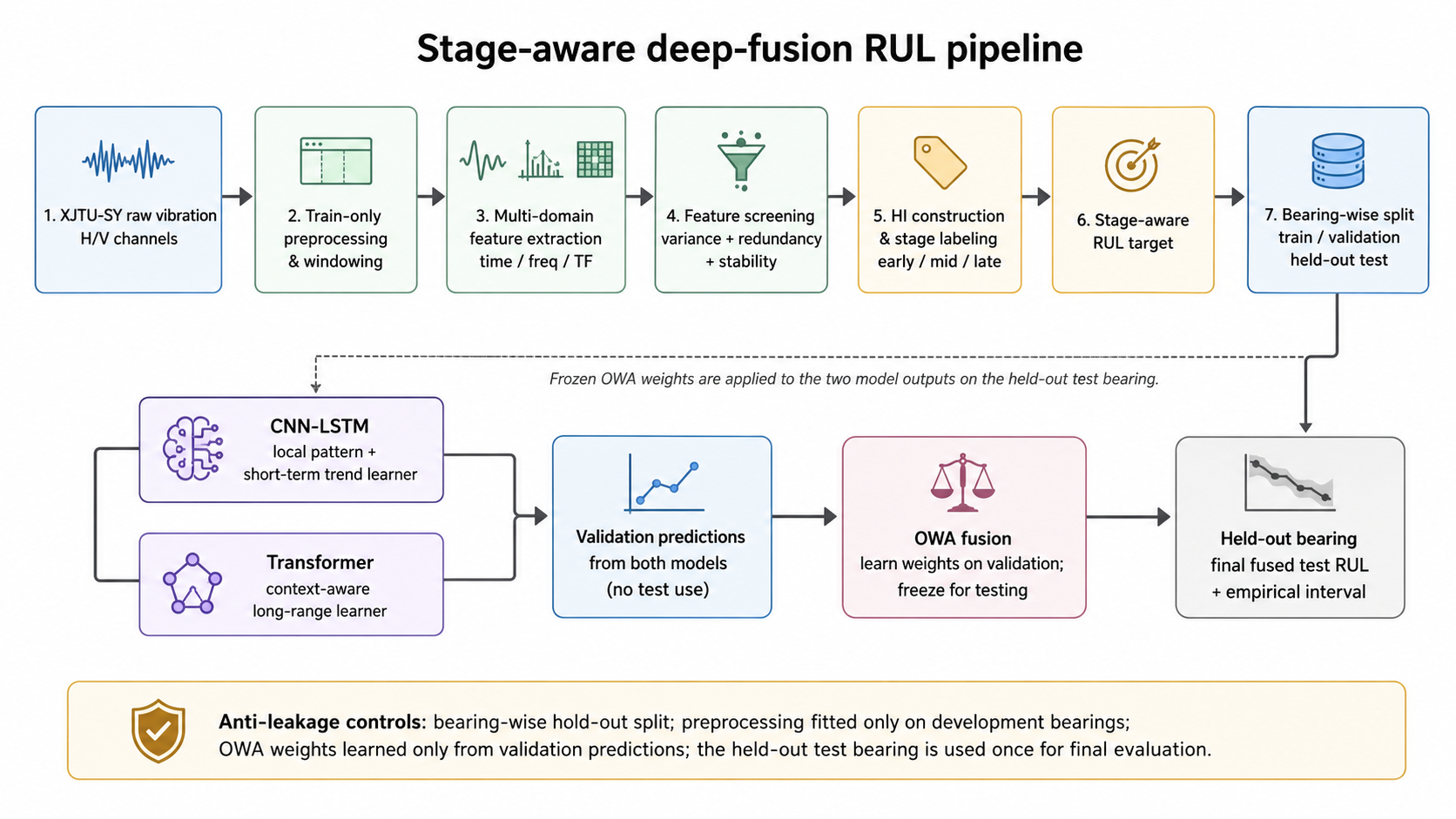}
    \caption{Study logic from raw vibration to evidence. Development bearings determine preprocessing, feature screening, networks, early stopping, and OWA weights. Complete trajectories are used retrospectively only to construct reference targets. XJTU-SY tests predictive learnability on held-out bearings; IMS tests target shape on a different rig and provides an exploratory cross-run target ablation.}
    \label{fig:pipeline}
\end{figure*}

The method first extracts multi-domain vibration descriptors using development data only, constructs an oriented principal-component health indicator, and repairs unsupervised stage labels into one chronological early--middle--late sequence. A continuous linear--quadratic--exponential curve is then fitted as a phenomenological degradation-state target. For prediction, a CNN--LSTM captures local sequential structure and a Transformer captures longer temporal context. Their scalar outputs are combined by an exact validation-constrained two-expert OWA rule \cite{yager1988owa}. The broader value of combining complementary sensing evidence is also illustrated in topology-aware hybrid Wi-Fi/BLE fingerprinting, where evidence-theoretic fusion integrates heterogeneous modalities \cite{mousaei2025topology}.

The evidence is deliberately hierarchical. The complete predictor is evaluated on XJTU-SY using Bearings~1--4 for development and Bearing~5 for final testing within each operating condition \cite{xjtuDataset,xjtuTutorial2019}. The target-shape hypothesis is assessed separately on documented IMS failed-bearing runs acquired on another rig \cite{imsDataset,imsBearingDataset2007}. The IMS comparison uses the best anchored linear approximation to the same health reference, not merely $1-u$, and reports information criteria together with a moving-block bootstrap. A small leave-one-run-out IMS experiment is retained as a negative-control style ablation: it tests whether a better target alone is sufficient under severe cross-run and cross-fault shift.

The contributions are:
\begin{enumerate}[leftmargin=*]
    \item a clear separation between \emph{target validity}, \emph{predictive accuracy}, and \emph{cross-domain transfer}, preventing these different claims from being conflated;
    \item a leakage-audited, stage-aware degradation-state target with explicit health-indicator orientation, chronological repair, continuity constraints, and comparison against a fitted linear baseline;
    \item a compact, auditable dual-scale predictor with branch-level ablation and closed-form validation-fitted OWA fusion; and
    \item a two-dataset evidence design that reports strong and weak results together, including the unresolved IMS directory mismatch, bootstrap uncertainty, and poor exploratory cross-run generalisation.
\end{enumerate}

The remainder of the paper defines the target and predictor, explains the evidence hierarchy, answers the three research questions, and discusses what the results do and do not support.

\section{Stage-Aware Target and Dual-Scale Predictor}
\label{sec:proposed_model}
The framework contains two linked but separately evaluated components: a retrospective target-construction procedure and a causal sequence predictor. The target is retrospective because a complete run-to-failure trajectory is needed to define its reference curve. The predictor is causal because each prediction uses only the current and preceding feature windows. This distinction is maintained throughout the methods and results.

The processing chain has six phases: (i) signal preparation and windowing, (ii) multi-domain feature extraction, (iii) development-only feature screening, (iv) health-indicator construction and chronological stage identification, (v) continuous stage-aware target fitting, and (vi) CNN--LSTM/Transformer prediction with validation-fitted OWA. No final-test bearing contributes to a fitted preprocessing statistic, feature threshold, network parameter, early-stopping decision, or fusion weight.

\begin{figure}[htbp]
	\centering
	\includegraphics[width=1\linewidth]{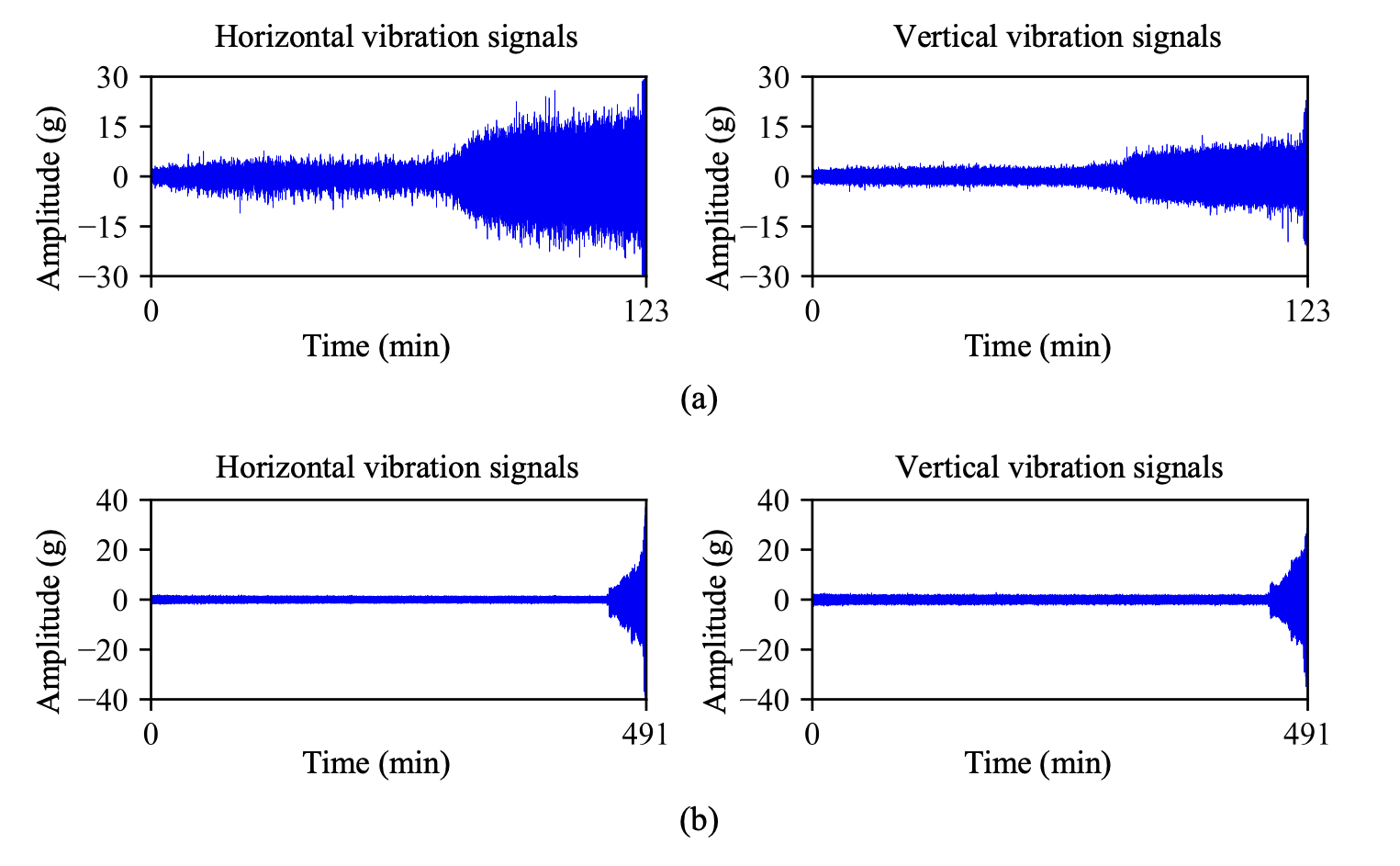}
	\caption[Raw vibration signals]{Example horizontal and vertical vibration signals from the XJTU-SY bearing dataset.}
	\label{fig:2}
\end{figure}

\subsection{Signal Preprocessing and Windowing}
\label{sec:preproc}
Let $x_{h}(t)$ and $x_{v}(t)$ denote raw horizontal and vertical vibration channels sampled at rate $f_s$. Each channel is de-trended and segmented with a sliding window of length $W$ samples and step $S$ (overlap $\eta=1-\tfrac{S}{W}$). Optional filtering is applied only with parameters fixed on development bearings, so that no information from the held-out test bearings determines preprocessing choices. Unless otherwise stated, $W$ and $S$ are selected to cover multiple shaft revolutions under each operating condition. The following normalisations are \emph{computed within the development bearings only} and then frozen:
\begin{enumerate}[leftmargin=*]
		\item \textbf{Condition-wise z-score:} for a windowed vector $\mathbf{x}$ from operating condition $c$, $\tilde{\mathbf{x}}=(\mathbf{x}-\mu_c)/\sigma_c$, with $\mu_c,\sigma_c$ computed only from the development bearings of the same condition and then reused for validation/test bearings.
		\item \textbf{Energy normalisation:} when features depend on absolute amplitude (e.g., RMS), we retain both the raw and the energy-normalised variant to preserve sensitivity to load changes.
	\end{enumerate}
To mitigate class imbalance across stages, we use stage-aware sampling in mini-batches, as described in the training strategy subsection.

\subsection{Feature Extraction Across Domains}
\label{sec:features}
For each window and channel, we compute a comprehensive descriptor bank spanning time, frequency, and time--frequency domains. We write $\mathcal{X}=\{x_h, x_v\}$ and form feature matrix $\mathbf{F}\in\mathbb{R}^{N\times d}$ by concatenation across channels.

\paragraph*{Time-domain descriptors.} Mean $\mu$, variance $\sigma^2$, standard deviation $\sigma$, RMS, median, inter-quartile range, skewness, kurtosis, zero-crossing rate, peak-to-peak, crest factor $c_f=\tfrac{\max|x|}{\mathrm{RMS}}$, shape factor $s_f=\tfrac{\mathrm{RMS}}{\tfrac{1}{n}\sum |x_i|}$, margin factor, impulse factor, and Hjorth parameters (activity, mobility, complexity).

\paragraph*{Frequency-domain descriptors.} From the Welch PSD $\hat{S}_{xx}(f)$ we compute spectral centroid $f_c=\frac{\int f \hat{S}_{xx}(f)\,df}{\int \hat{S}_{xx}(f)\,df}$, spectral bandwidth, spectral entropy $-\sum p_f\log p_f$ with $p_f\propto \hat{S}_{xx}(f)$, spectral kurtosis, dominant frequencies $f_{(1:3)}$ and their amplitudes, and band powers $\int_{B_k}\hat{S}_{xx}(f)\,df$ for bands $\{B_k\}$ that cover known bearing-fault harmonics as well as broad bands.

\paragraph*{Time–frequency descriptors.} Short-time Fourier transform (STFT) statistics (band-limited energies and entropy across time) and wavelet-packet energies aggregated per sub-band. To avoid leakage, the choice of mother wavelet/band grid is fixed on the training set and applied unchanged to the test set.

\paragraph*{Cross-channel fusion features.} We include channel-pair features such as
\begin{equation}
\mathrm{RMS}_{\mathrm{HV}}=\sqrt{\mathrm{RMS}_h^2+\mathrm{RMS}_v^2},
\end{equation}
and the within-window correlation $\rho(x_h,x_v)$, which often become monotone as wear progresses.

\subsection{Development-Only Feature Screening}
\label{sec:featselect}
Feature screening is performed using development bearings only. The aim is to remove descriptors that are numerically uninformative, duplicate another descriptor, or vary erratically across runs. Let $\mathbf{F}\in\mathbb{R}^{N\times d}$ denote the development feature matrix.

\begin{enumerate}[leftmargin=*]
    \item \textbf{Low-variance filter:} discard $f_j$ when
    \begin{equation}
        \operatorname{Var}(f_j)<\tau_{\mathrm{var}},
    \end{equation}
    where $\tau_{\mathrm{var}}$ is computed from development data only.

    \item \textbf{Redundancy filter:} if $|\rho(f_j,f_k)|>\tau_\rho$, retain the descriptor with the larger temporal-separation score
    \begin{equation}
        S_j=\frac{(\mu_{j,\mathcal{D}}-\mu_{j,\mathcal{H}})^2}
        {\sigma^2_{j,\mathcal{D}}+\sigma^2_{j,\mathcal{H}}+\varepsilon}.
    \end{equation}
    Here $\mathcal{H}$ and $\mathcal{D}$ are fixed early- and late-life temporal tails from the \emph{development} runs. They are proxy groups used only for screening and are defined before the subsequent $k$-means stage labels; this avoids circularly selecting features with labels produced by the same clustering procedure.

    \item \textbf{Robust stability filter:} for each run, compute
    \begin{equation}
        C^{\mathrm{rob}}_{v,j}=\frac{1.4826\,\operatorname{MAD}(f_j)}{|\operatorname{median}(f_j)|+\varepsilon}.
    \end{equation}
    A descriptor is rejected when this robust relative-dispersion measure exceeds its development-set threshold. The absolute value and $\varepsilon$ prevent the undefined or misleading behaviour of the conventional $\sigma/\mu$ coefficient when a feature mean is zero or negative.
\end{enumerate}

\begin{figure}[htbp]
    \centering
    \includegraphics[width=1\linewidth]{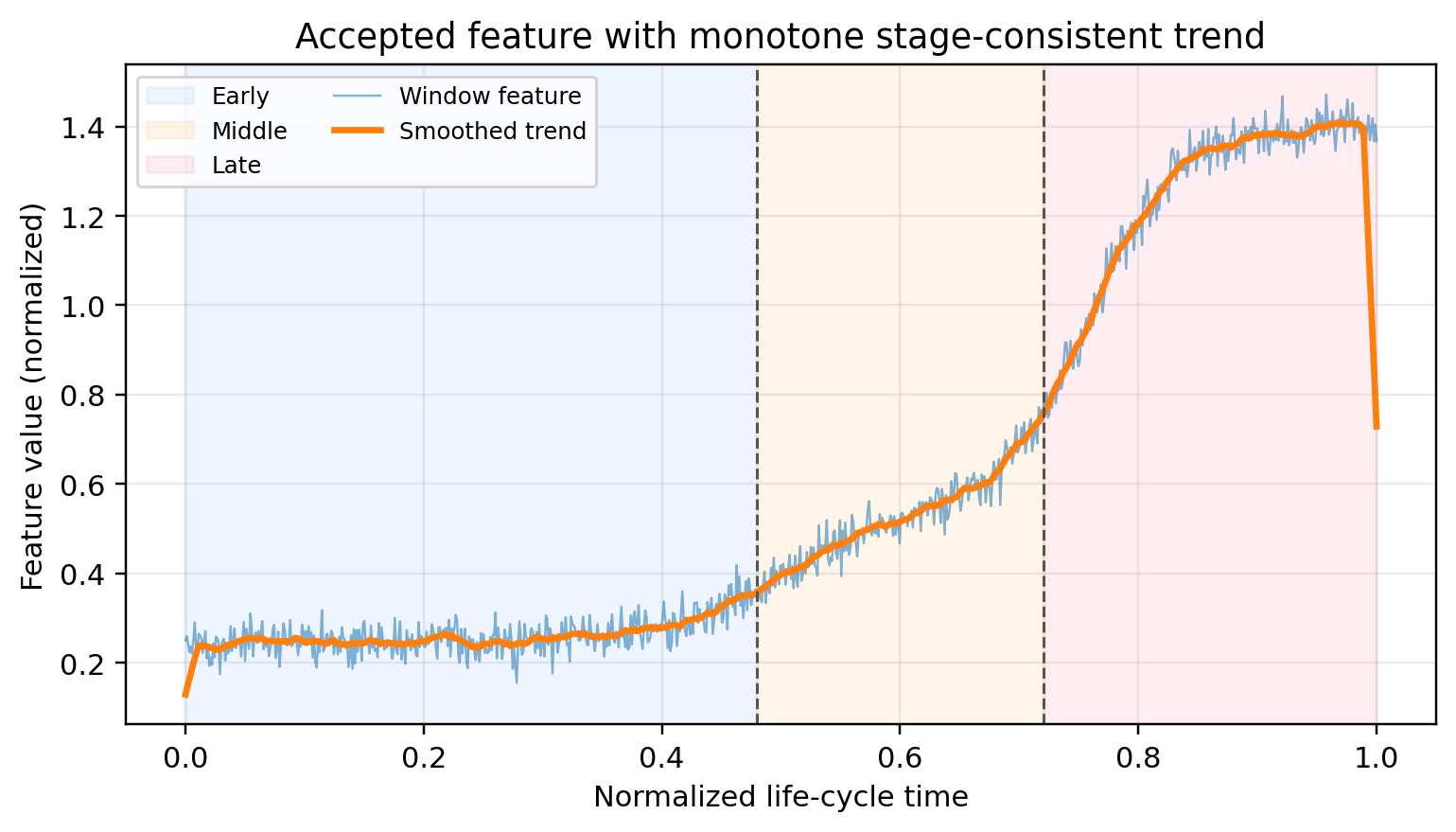}
    \caption[Accepted monotone feature]{Example retained feature with a monotone, stage-consistent trend after smoothing.}
    \label{fig:3}
\end{figure}

The resulting matrix $\mathbf{F}^{\star}\in\mathbb{R}^{N\times d^{\star}}$, with $d^{\star}\ll d$, is passed to target construction and sequence modelling. Thresholds, retained features, and the fitted scaling objects belong to the model and must be exported with the final implementation. Figure~\ref{fig:3} illustrates a retained descriptor with a stage-consistent trend.

\subsection{Life-Stage Identification Using K-Means Clustering}
\label{sec:staging}
The selected descriptors are used retrospectively to identify three degradation stages for each complete run-to-failure trajectory. This operation constructs supervision; it is not an online change-point detector.

\paragraph*{Monotonicity ranking.} For feature $f_j$ observed over a bearing run of length $T$, define
\begin{equation}
    M_j=\frac{|f_j(T)-f_j(1)|}{\sum_{t=2}^{T}|f_j(t)-f_j(t-1)|+\varepsilon}.
\end{equation}
The score approaches one for a smoothly monotone trajectory. Spearman correlation $\rho^{\mathrm{S}}(f_j,t)$ is used as a second trend criterion.

\paragraph*{Health-indicator construction and orientation.} The $q$ highest-ranked descriptors form a low-dimensional health indicator through the first principal component,
\begin{equation}
    h_b(t)=\operatorname{PCA}_1\!\left(\{f_j(t)\}_{j=1}^{q}\right).
\end{equation}
Because the sign of a principal component is arbitrary, it is oriented explicitly as
\begin{equation}
    h_b(t)\leftarrow \operatorname{sign}\!\left(\rho^{\mathrm{S}}(h_b,t)\right)h_b(t),
\end{equation}
so that larger values correspond to later degradation. LOWESS or median smoothing is then applied to reduce window-level jitter without changing temporal order.

\paragraph*{Stage clustering and chronological repair.} Three one-dimensional centroids are estimated from the smoothed indicator:
\begin{equation}
    \min_{\{\mathcal{C}_k\}_{k=1}^{3}}
    \sum_{k=1}^{3}\sum_{t\in\mathcal{C}_k}
    \left|h_b(t)-\mu_k\right|^2.
\end{equation}
The clusters are ordered by their median time index and converted into contiguous stages $\mathcal{S}_1\prec\mathcal{S}_2\prec\mathcal{S}_3$. Short isolated segments are merged with a neighbouring stage, and an audit enforces the sequence $1\rightarrow2\rightarrow3$. This pragmatic procedure captures common stage structure but is not a mechanistic fault-physics model.

\begin{figure}[htbp]
    \centering
    \includegraphics[width=1\linewidth]{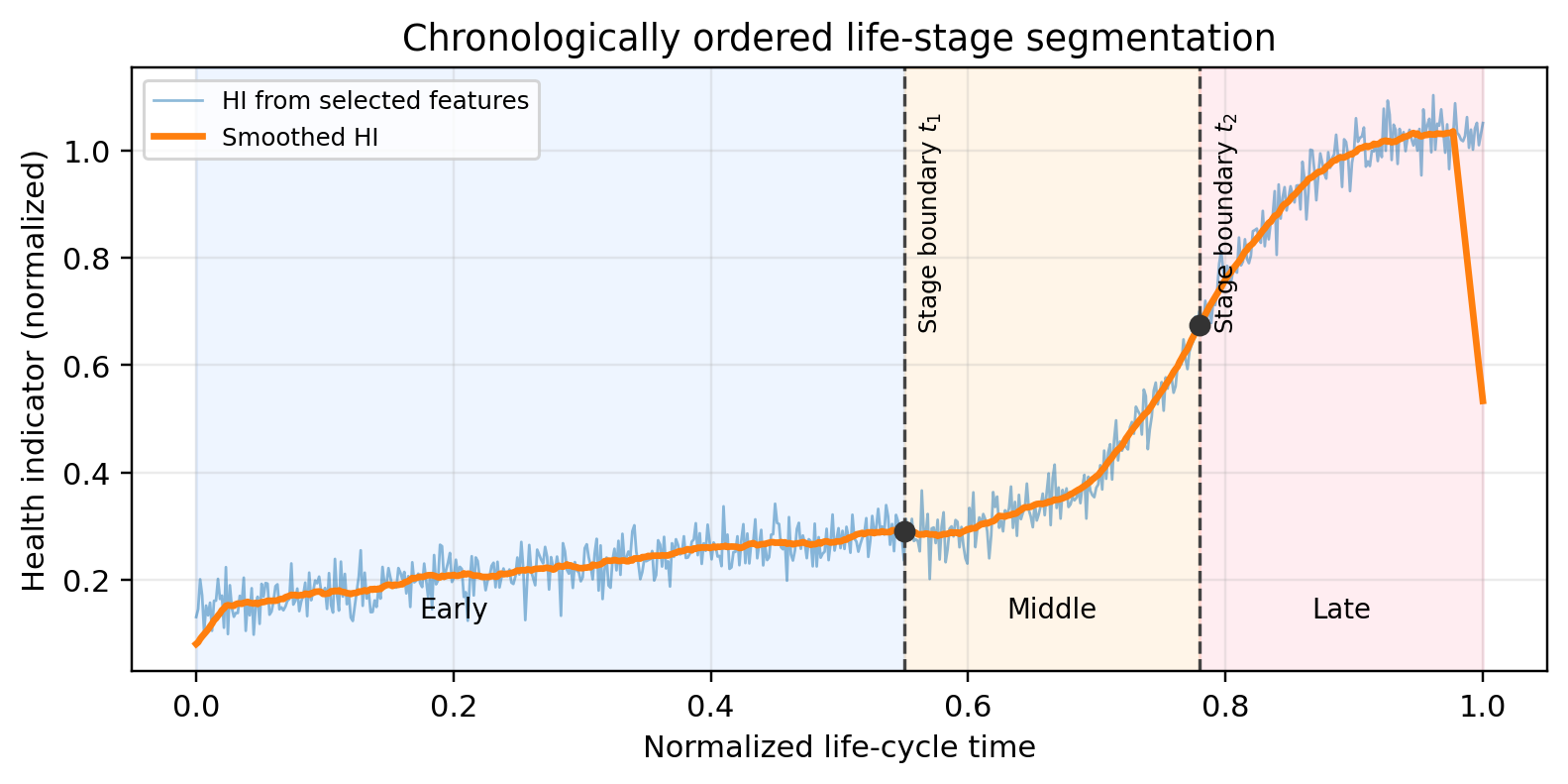}
    \caption[Chronological life-stage segmentation]{Chronologically ordered life-stage segmentation obtained from the oriented, smoothed health indicator.}
    \label{fig:9}
\end{figure}

For development bearings, the complete trajectory is used only to create training labels. For held-out bearings, the same retrospective procedure may be used after failure solely to create reference curves for scoring; those reference labels are not inputs to preprocessing, model fitting, early stopping, or OWA estimation. Figure~\ref{fig:9} shows one example.

\subsection{Stage-Wise Normalised Remaining-Life Target}
\label{sec:piecewise}
The oriented health indicator increases with degradation, whereas remaining life decreases. We therefore normalise the indicator within a complete bearing run,
\begin{equation}
    \widetilde{h}_b(u)=\frac{h_b(u)-\min h_b}{\max h_b-\min h_b+\varepsilon},
\end{equation}
and define the complementary degradation reference $z_b(u)=1-\widetilde{h}_b(u)$, where $u=t/T_b\in[0,1]$ is normalised life-cycle time. The regression target is a continuous piecewise approximation to $z_b$, not a direct physical measurement of time remaining:
\begin{equation}\label{eq:piecewiseRUL}
R_b(u)=
\begin{cases}
1-\alpha u, & u\in\mathcal{S}_1,\\[4pt]
1-\alpha u-\beta(u-u_1)^2, & u\in\mathcal{S}_2,\\[4pt]
R_b(u_2)\exp[-\gamma(u-u_2)], & u\in\mathcal{S}_3.
\end{cases}
\end{equation}
The transition locations $u_1$ and $u_2$ are obtained from the repaired stage sequence. Parameters $\alpha,\beta,\gamma\ge0$ are fitted by least squares to $z_b(u)$ with $C^0$ continuity, $R_b(0)=1$, and $R_b(1)\approx0$. A weak optional slope penalty at the second boundary is
\begin{equation}
\lambda_2\left[-\alpha-2\beta(u_2-u_1)+\gamma R_b(u_2)\right]^2.
\end{equation}
Finally, $R_b$ is clipped to $[0,1]$. This construction encodes a mild early decline, accelerated middle degradation, and a steep terminal decline. It is degradation-informed and physically motivated, but it should not be described as a universal physics law.

\begin{figure}[htbp]
    \centering
    \includegraphics[width=1\linewidth]{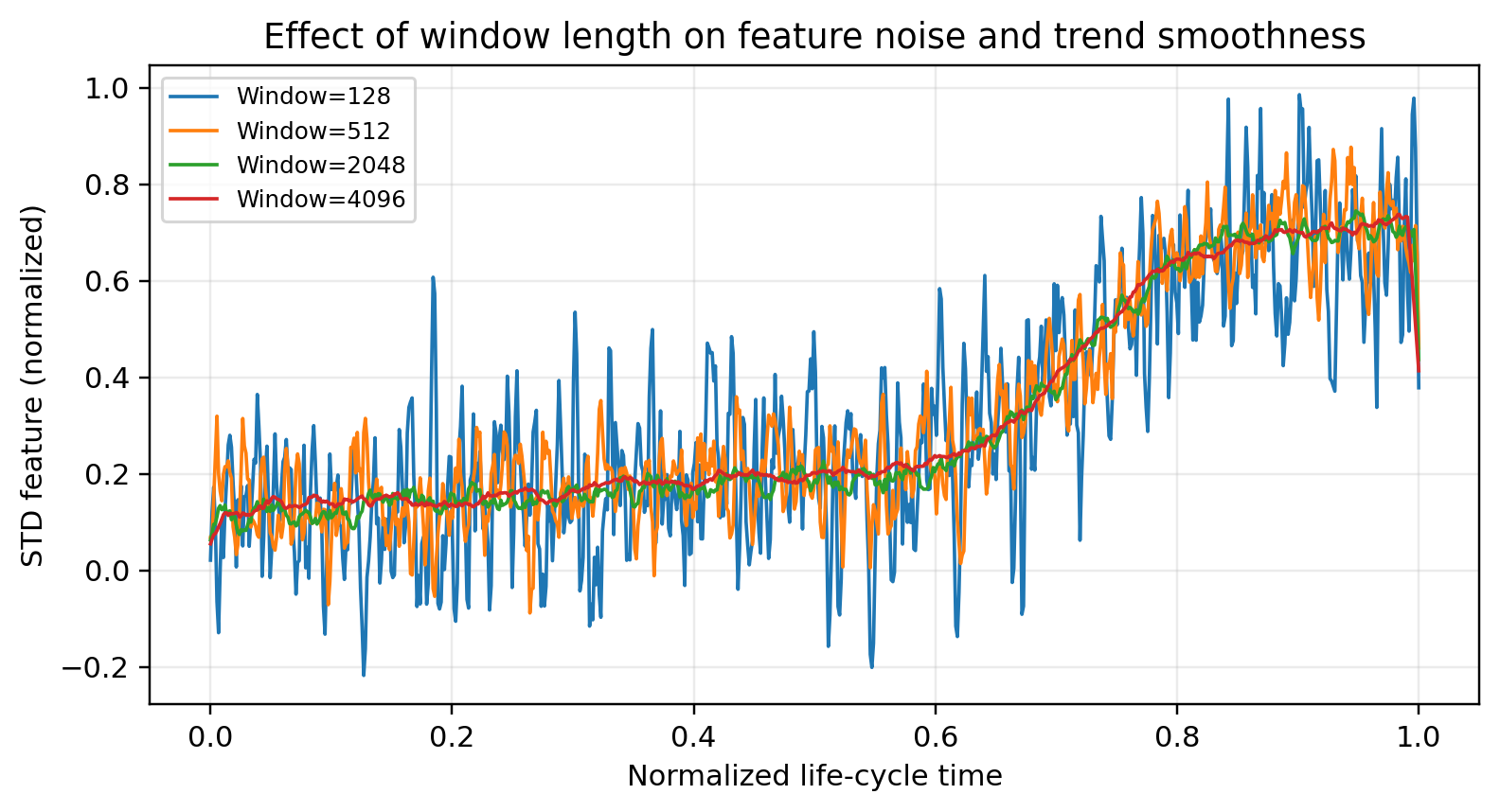}
    \caption[Window-size effect on feature stability]{Illustration of the trade-off between local feature noise and trend smoothness for different window lengths.}
    \label{fig:10}
\end{figure}

Longer windows can suppress high-frequency feature jitter at the expense of temporal resolution. Figure~\ref{fig:10} illustrates this bias--variance trade-off; the final window length must be fixed using development data and reported with the experiment configuration.

\begin{figure}[htbp]
    \centering
    \includegraphics[width=1\linewidth]{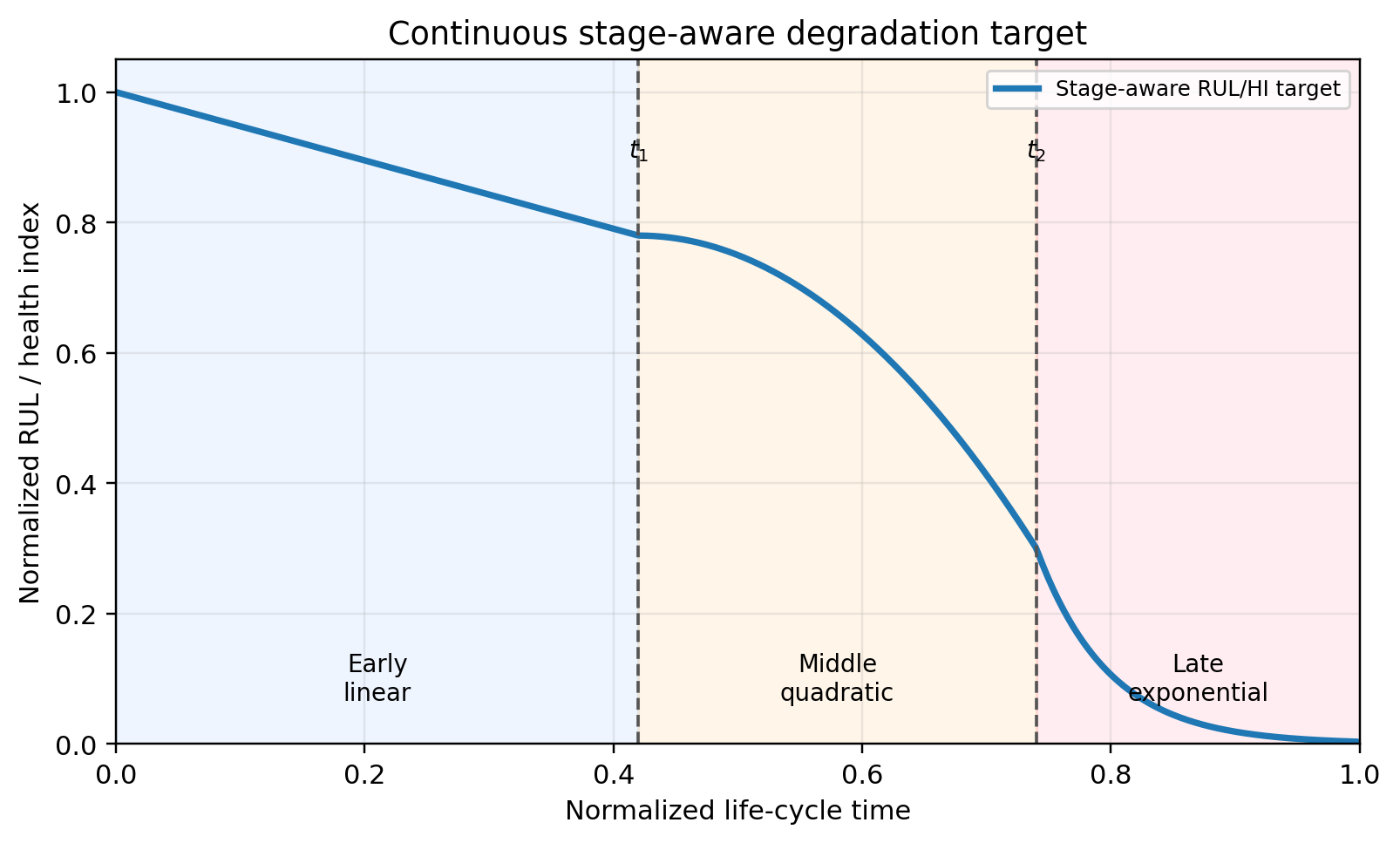}
    \caption[Stage-aware RUL target]{Continuous degradation-informed target with linear, quadratic, and exponential segments.}
    \label{fig:1}
\end{figure}

\subsection{Linear Baselines and Evidence for Stage Dependence}
\label{sec:linear_baselines}
The stage-aware curve is evaluated against two linear references. The first is the conventional clock target
\begin{equation}
R_{\mathrm{clock}}(u)=1-u.
\end{equation}
Because this fixed-slope target may be disadvantaged simply by scale, a stronger baseline is fitted to the same vibration-derived reference $z_b(u)$:
\begin{align}
R_{\mathrm{lin}}(u;a)&=\operatorname{clip}_{[0,1]}(1-au),\\
a^{\star}&=\arg\min_{0\le a\le2.5}\frac{1}{T_b}\sum_{t=1}^{T_b}\left[R_{\mathrm{lin}}(u_t;a)-z_b(u_t)\right]^2.
\end{align}
The main target-shape comparison is therefore between $R_{\mathrm{lin}}(u;a^{\star})$ and the three-stage curve in Eq.~\eqref{eq:piecewiseRUL}, rather than only against $1-u$.

For each complete run, RMSE, MAE, $R^2$, AIC, and BIC are computed against $z_b$. Positive
\begin{equation}
\Delta\mathrm{BIC}=\mathrm{BIC}_{\mathrm{linear}}-\mathrm{BIC}_{\mathrm{stage}}
\end{equation}
favours the stage-aware curve. To avoid understating its flexibility, the conservative BIC calculation assigns five effective parameters to the stage model: $\alpha$, $\beta$, $\gamma$, $u_1$, and $u_2$, even though the transition locations are obtained from the clustering procedure. Temporal dependence is addressed using a moving-block bootstrap on the squared-error gain
\begin{equation}
\delta_t=\left[z_b(u_t)-R_{\mathrm{lin}}(u_t)\right]^2-
\left[z_b(u_t)-R_b(u_t)\right]^2,
\end{equation}
where positive mean $\delta_t$ favours the stage-aware representation. Blocks contain approximately 3\% of each run, and the reported experiment uses 300 bootstrap repetitions. These tests evaluate descriptive adequacy of the target representation; they do not convert the retrospective surrogate into a direct physical measurement of time remaining.

\subsection{Training Strategy}
\label{sec:training}
\subsubsection*{Bearing-wise test separation}
For each operating condition, Bearings~1--4 are assigned to development and Bearing~5 is reserved for final evaluation (Table~\ref{tab:dataset_config}). Consequently, \texttt{Bearing1\_5}, \texttt{Bearing2\_5}, and \texttt{Bearing3\_5} do not contribute to fitted normalisation statistics, feature-screening thresholds, network parameters, early stopping, hyperparameter choices, or OWA weights.

\subsubsection*{Validation and early stopping}
Within the development bearings, 80\% of windows are used for parameter learning and 20\% for validation with \texttt{random\_state}=42. Validation loss controls early stopping and supplies the predictions used to fit OWA weights. Because neighbouring windows from the same physical bearing are highly correlated, this random window-level validation split is optimistic and is \emph{not} treated as independent evidence of generalisation. The held-out-bearing test is the main generalisation result. A stronger future protocol should apply group-wise or blocked validation across multiple bearing assignments and random seeds.

\subsubsection*{Causality and retrospective labels}
The stage-aware target is generated from a complete run-to-failure record and therefore uses future observations relative to an individual training time point. This is permissible only as retrospective supervision. Predictor sequences are formed from the current and preceding feature windows; no future signal window is supplied as a model input. At evaluation, full test-run information is used only to construct the reference target after the run has ended, never to tune the predictor or fusion rule.

\subsubsection*{Regularisation}
Mini-batches are sampled to reduce early/middle/late imbalance. The networks use $\ell_2$ weight decay, dropout, gradient clipping, and early stopping. An optional monotonicity penalty can discourage increasing remaining-life predictions within a sequence:
\begin{equation}
\mathcal{L}_{\mathrm{mono}}=\lambda_{\mathrm{m}}\sum_{t>1}\max\left(0,\hat{R}(t)-\hat{R}(t-1)\right).
\end{equation}
The benchmark values reported in this paper use mean-squared error as the base loss.
\subsection{Validation-Optimised OWA Decision Fusion}
The two branch predictions are combined with the Ordered Weighted Averaging operator introduced by Yager \cite{yager1988owa}. For sample $i$, let $p_{(1),i}\le p_{(2),i}$ be the sorted CNN--LSTM and Transformer predictions. The fused prediction is
\begin{align}
    \widehat{R}_{i}&=w_1p_{(1),i}+w_2p_{(2),i},\\
    w_1,w_2&\ge0,\qquad w_1+w_2=1.
\end{align}
The sorting is sample-specific, but the weights are global parameters learned once on validation data and frozen for all test bearings. Thus, the method is rank-adaptive, not a sample-conditioned gating network. A large $w_1$ favours the lower, more conservative estimate; a large $w_2$ favours the higher estimate.

For two experts, the constrained least-squares solution can be obtained exactly without solving an unconstrained two-parameter problem and subsequently projecting it. Let $d_i=p_{(1),i}-p_{(2),i}$. Since $w_2=1-w_1$,
\begin{equation}
    \widehat{R}_i=p_{(2),i}+w_1d_i.
\end{equation}
The validation-optimal weight is
\begin{align}\label{eq:owa_closed}
    w_1^{\star}&=\operatorname{clip}_{[0,1]}\!\left(
    \frac{\sum_{i=1}^{m}d_i\left(y_i-p_{(2),i}\right)}
    {\sum_{i=1}^{m}d_i^2+\varepsilon}\right),\\
    w_2^{\star}&=1-w_1^{\star}.
\end{align}
If both branches make identical validation predictions, equal weights are used. The learned pair is then frozen. Reporting the numerical weights and their variability under resampling is recommended because unstable weights would indicate that the apparent complementarity is not robust.

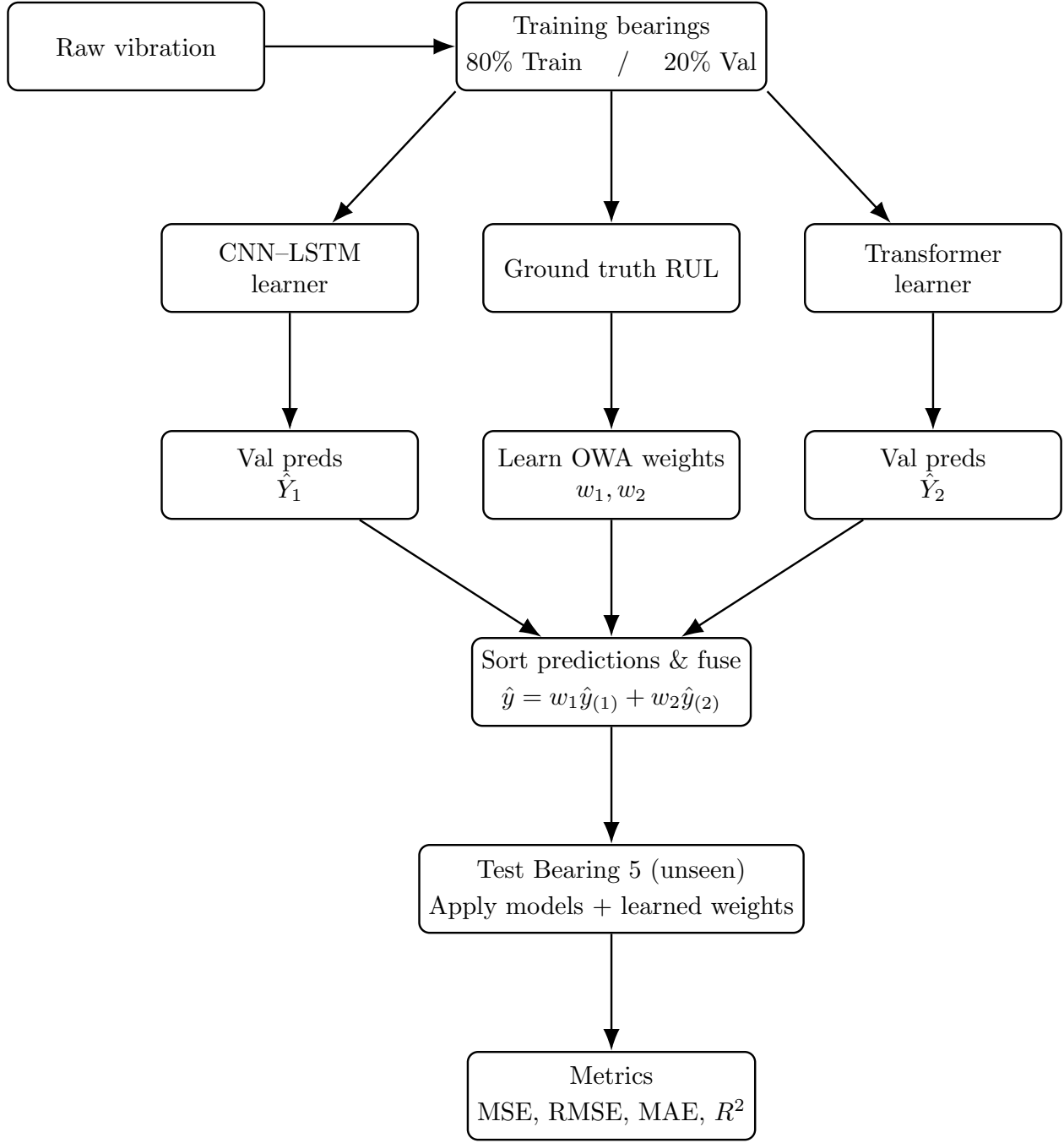
\begin{figure}[ht]
	\centering
	\resizebox{\columnwidth}{!}{%
		\begin{tikzpicture}[
			node distance=1.6cm and 2.6cm,
			font=\small
			]
			
			% --- Row 1: Input & data split ---
			\node[block] (raw) {Raw vibration};
			\node[block, right=of raw] (split) {Training bearings \\[2pt]
					80\% Train \quad / \quad 20\% Val};
			
			% --- Row 2: Learners ---
			\node[block, below left=1.8cm and 0.5cm of split] (cnn) {CNN--LSTM \\ learner};
			\node[block, below right=1.8cm and 0.5cm of split] (trs) {Transformer \\ learner};
			
			% --- Row 3: Dev/Val predictions ---
			\node[block, below=of cnn] (y1) {Val preds \\ $\hat{Y}_1$};
			\node[block, below=of trs] (y2) {Val preds \\ $\hat{Y}_2$};
			
			% --- OWA weights & GT RUL ---
			\node[block, below=1.8cm of split] (gt) {Ground truth RUL};
			\node[block, below=of gt] (owa) {Learn OWA weights \\ $w_1, w_2$};
			
			% --- Fusion ---
			\node[block, below=of owa] (fusion) {Sort predictions \& fuse \\[2pt]
					$\hat{y} = w_1 \hat{y}_{(1)} + w_2 \hat{y}_{(2)}$};
			
			% --- Test and metrics ---
			\node[block, below=of fusion] (test) {Test Bearing 5 (unseen) \\[2pt]
				Apply models + learned weights};
			\node[block, below=of test] (metrics) {Metrics \\[2pt]
				MSE, RMSE, MAE, $R^2$};
			
			% --- Connections ---
			\draw[line] (raw) -- (split);
			
			\draw[line] (split.south west) -- (cnn);
			\draw[line] (split.south east) -- (trs);
			
			\draw[line] (cnn) -- (y1);
			\draw[line] (trs) -- (y2);
			
			\draw[line] (split) -- (gt);
			\draw[line] (gt) -- (owa);
			
			\draw[line] (y1) -- (fusion);
			\draw[line] (y2) -- (fusion);
			\draw[line] (owa) -- (fusion);
			
			\draw[line] (fusion) -- (test);
			\draw[line] (test) -- (metrics);
			
		\end{tikzpicture}%
	}
	\caption[Diagram of fusing two models using OWA on validation data and testing the fused model on test data]{Validation-based OWA fusion procedure for combining the CNN-LSTM and Transformer predictors before testing on held-out bearings.}
	\label{fig:101}
\end{figure}

\subsection{Network Architectures and Hyperparameters}
\label{sec:nets}
In the following, we detail the architectures of the CNN--LSTM and Transformer models used for Remaining Useful Life (RUL) estimation. We keep the base designs deliberately compact and make the implementation choices explicit (activations, initialisation, regularisation) to aid reproducibility and future deployment.

\paragraph*{CNN--LSTM pathway.} A shallow Conv1D front-end is applied to sequences of consecutive feature windows and captures local temporal changes; LSTM modelling and two dense layers follow. Rectified linear units are used throughout except for the output (softplus to ensure non-negativity). Batch normalisation after the convolution stabilises training; dropout is applied before the dense block; the exact dropout rate must be reported from the final experiment configuration.

	\paragraph*{Transformer pathway.} Each input sequence contains $L$ consecutive feature windows, so self-attention operates over time rather than over a single static descriptor vector. A linear embedding maps descriptors to $d_\mathrm{model}{=}32$, positional encodings preserve temporal order, and a Transformer block with $h{=}2$ heads and feed-forward width $64$ models longer-range temporal context. We use pre-norm residual connections and dropout in the multi-head attention and feed-forward sub-layers. GlobalAveragePooling1D aggregates the temporal dimension before the dense head.

\begin{table}[htbp]
		\centering
		\caption{Model Architectures}
		\footnotesize
		\begin{tabularx}{\columnwidth}{@{}p{0.34\columnwidth}X@{}}
			\toprule
			\textbf{Layer Type} & \textbf{CNN-LSTM Model} \\
			\midrule
			Input & Sequence tensor $(L,d^{\star})$ \\
			Feature extraction & Conv1D, 16 filters, kernel size 3, ReLU; MaxPooling1D, pool size 2 \\
			Temporal processing & LSTM, 32 units \\
			Output processing & Dense, 16 units, ReLU \\
			Output layer & Dense, 1 unit, softplus; predictions clipped to $[0,1]$ for evaluation \\
			\midrule
			\textbf{Layer Type} & \textbf{Transformer Model} \\
			\midrule
			Input & Sequence tensor $(L,d^{\star})$ \\
			Feature extraction & Dense embedding, 32 units, with positional encoding \\
			Temporal processing & Transformer block, embed\_dim=32, num\_heads=2, ff\_dim=64; GlobalAveragePooling1D \\
			Output processing & Dense, 16 units, ReLU \\
			Output layer & Dense, 1 unit, softplus; predictions clipped to $[0,1]$ for evaluation \\
			\bottomrule
		\end{tabularx}
	\end{table}

Both models were trained using the Adam optimiser with the following hyperparameters and training configurations.

\begin{table}[htbp]
    \centering
    \caption{Optimiser and hyperparameters}
    \footnotesize
    \begin{tabularx}{\columnwidth}{@{}p{0.31\columnwidth}X@{}}
        \toprule
        \textbf{Parameter} & \textbf{Value} \\
        \midrule
        Optimiser & Adam \\
        Learning rate & 0.001 \\
        Loss function & Mean-squared error for the reported benchmark \\
        Reported metric & Mean-absolute error \\
        Maximum epochs & 100 \\
        Batch size & 32 \\
        Validation split & 20\% of development windows \\
        Early stopping & Patience of 10 epochs on validation loss \\
        \bottomrule
    \end{tabularx}
\end{table}

\paragraph*{Implementation notes.} We initialise weights with He/Kaiming initialisation for ReLU layers and Xavier for linear/attention blocks; biases are zero-initialised. Gradients are clipped to a max-norm of~1.0. Inputs and targets are cached in contiguous memory to avoid I/O bottlenecks; training uses mixed precision where available.

\section{Experimental Design and Results}
\label{sec:experiments}
The experiments are organised by claim rather than by dataset. RQ1 concerns descriptive adequacy of the target on an external rig. RQ2 concerns held-out-bearing prediction. RQ3 concerns the incremental contribution of fusion and the limits of target choice under distribution shift. Table~\ref{tab:evidence_hierarchy} specifies the unit of independence and the permitted interpretation for each experiment.

\begin{table}[!t]
\centering
\scriptsize
\caption{Research questions, evidence units, and claim boundaries.}
\label{tab:evidence_hierarchy}
\begin{tabularx}{\textwidth}{@{}p{0.08\textwidth}p{0.22\textwidth}p{0.19\textwidth}p{0.20\textwidth}X@{}}
\toprule
\textbf{RQ} & \textbf{Experiment} & \textbf{Independent unit} & \textbf{Primary comparison} & \textbf{Permitted claim} \\
\midrule
RQ1 & IMS external-dataset target-shape assessment & Complete documented failed-bearing trajectory & Stage-aware curve versus best anchored linear fit to the same HI reference & Descriptive support for stage-dependent degradation-state labels; not online RUL prediction \\
RQ2 & XJTU-SY bearing-wise predictive benchmark & Bearing excluded from all fitted development steps & CNN--LSTM, Transformer, and frozen OWA fusion & Predictive learnability of the retrospective target on the predefined hold-out \\
RQ3 & XJTU branch ablation and exploratory IMS leave-one-run-out target ablation & Held-out bearing/run & Single branches versus OWA; clock-linear versus stage-aware supervision & Incremental fusion value and evidence that target design alone does not solve domain shift \\
\bottomrule
\end{tabularx}
\end{table}

\subsection{Datasets and Evaluation Roles}
\label{sec:datasets}
The two datasets are used for complementary purposes rather than pooled into one training corpus. XJTU-SY provides 15 complete trajectories under three speed/load combinations and supports the principal bearing-wise predictive benchmark. IMS contains three documented test-to-failure experiments from a different bearing type, rig, sampling rate, channel arrangement, and acquisition interval; it is used primarily to test whether the nonlinear target construction transfers beyond XJTU-SY. Table~\ref{tab:dataset_overview} summarises the acquisition differences from the official dataset documentation \cite{xjtuTutorial2019,imsBearingDataset2007}.

\begin{table}[!t]
\centering
\scriptsize
\caption{Acquisition characteristics and evaluation roles of the two bearing datasets.}
\label{tab:dataset_overview}
\begin{tabularx}{\textwidth}{@{}p{0.12\textwidth}p{0.17\textwidth}p{0.19\textwidth}p{0.20\textwidth}X@{}}
\toprule
\textbf{Dataset} & \textbf{Rig and operating conditions} & \textbf{Sampling} & \textbf{Run-to-failure content} & \textbf{Role in this study} \\
\midrule
XJTU-SY \cite{xjtuDataset,xjtuTutorial2019} & LDK UER204 bearings; 2100/2250/2400~r/min under 12/11/10~kN & Two accelerometers (horizontal/vertical); 25.6~kHz; 32,768 points (1.28~s) every 1~min & 15 bearings, five per condition; tests stop after severe vibration relative to the normal stage & Complete CNN--LSTM/Transformer/OWA prediction benchmark with Bearings~1--4 for development and Bearing~5 for final testing in each condition \\
IMS \cite{imsDataset,imsBearingDataset2007} & Rexnord ZA-2115 bearings; 2000~r/min under a 6000-lb radial load & 20~kHz; 20,480 points (1~s); nominal 10-min interval; Set~1 has two channels per bearing and Sets~2/3 one channel per bearing & Three documented experiments with 2,156, 984, and 4,448 snapshots; documented inner-race, roller-element, and outer-race failures & External-dataset target-shape assessment on documented failed bearings plus an explicitly exploratory leave-one-run-out target-training ablation \\
\bottomrule
\end{tabularx}
\end{table}

\subsection{RQ1: External-Dataset Target-Shape Assessment}
\label{sec:ims_validation}
The IMS bearing data were generated on a four-bearing test rig operating at 2000~r/min under a 6000-lb radial load. Each timestamped file contains a 1-s vibration snapshot with 20,480 samples at 20~kHz. Set~1 contains eight channels (two per bearing), while Sets~2 and~3 contain one channel per bearing \cite{imsDataset,imsBearingDataset2007}. The documented failures used in the confirmatory analysis are Set~1 Bearing~3 (inner-race defect), Set~1 Bearing~4 (roller-element defect), and Set~2 Bearing~1 (outer-race failure). The official documentation also identifies Set~3 Bearing~3 as an outer-race failure, but the locally available directory does not match the documented Set~3 file count or end date and is therefore kept exploratory.

A directory and file-count audit was performed before feature extraction (Table~\ref{tab:ims_audit}). Sets~1 and~2 exactly match the official documentation. The available directory labelled \texttt{3rd\_test/4th\_test/txt} contains 6,324 files and ends on 18~April~2004, whereas the supplied README describes 4,448 Set~3 files ending on 4~April~2004. Results from this extended directory are retained for completeness but are excluded from the confirmatory three-run aggregate and marked exploratory.

\begin{table}[!t]
\centering
\scriptsize
\caption{IMS data audit. The first two sets match the supplied documentation; the extended directory does not match the documented Set~3 count or end date.}
\label{tab:ims_audit}
\begin{tabularx}{\textwidth}{@{}lclllX@{}}
\toprule
\textbf{Data source} & \textbf{Files} & \textbf{Channels} & \textbf{First timestamp} & \textbf{Last timestamp} & \textbf{Use in this study} \\
\midrule
IMS Set~1 & 2,156 & 8 & 22 Oct 2003 & 25 Nov 2003 & Confirmatory; Bearings~3 and~4 \\
IMS Set~2 & 984 & 4 & 12 Feb 2004 & 19 Feb 2004 & Confirmatory; Bearing~1 \\
Extended \texttt{4th\_test/txt} directory & 6,324 & 4 & 4 Mar 2004 & 18 Apr 2004 & Exploratory only; mismatch with documented Set~3 (4,448 files, ending 4 Apr 2004) \\
\bottomrule
\end{tabularx}
\end{table}

For each failed-bearing trajectory, time-, frequency-, wavelet-, and channel-fusion descriptors were extracted from every snapshot. Ten trend-consistent, non-redundant features were used to form the oriented PCA health indicator. The smoothing fraction was 0.035, the minimum stage fraction was 0.06, the wavelet was db4 at level~3, and the redundancy threshold was 0.98. The stage-aware curve was compared with the best anchored linear fit defined above. Conservative BIC treats the stage locations as fitted degrees of freedom, and the moving-block bootstrap uses 300 repetitions with block length approximately 3\% of each trajectory.

\begin{table}[!t]
\centering
\scriptsize
\caption{Linear versus stage-aware approximation of the IMS vibration-derived remaining-life reference. Positive RMSE/MAE gains and positive $\Delta$BIC favour the stage-aware representation. The final row is exploratory because of the audit mismatch in Table~\ref{tab:ims_audit}.}
\label{tab:ims_nonlinearity}
\resizebox{\textwidth}{!}{%
\begin{tabular}{llccccccccc}
\toprule
\textbf{Run} & \textbf{Fault} & $u_1$ & $u_2$ & \textbf{Lin. RMSE} & \textbf{Stage RMSE} & \textbf{RMSE gain} & \textbf{Lin. MAE} & \textbf{Stage MAE} & $\boldsymbol{\Delta}$\textbf{BIC} & $P(\overline{\delta}>0)$ \\
\midrule
IMS1-B3 & Inner race & 0.074 & 0.845 & 0.0972 & 0.0887 & 8.8\% & 0.0740 & 0.0717 & 368.1 & 0.49 \\
IMS1-B4 & Roller element & 0.077 & 0.748 & 0.1094 & 0.1054 & 3.6\% & 0.0922 & 0.0824 & 128.8 & 0.81 \\
IMS2-B1 & Outer race & 0.584 & 0.842 & 0.1356 & 0.1109 & 18.2\% & 0.1064 & 0.0733 & 367.2 & 0.83 \\
Extended-B3$^{\ast}$ & Label inherited from directory & 0.880 & 0.940 & 0.1113 & 0.0485 & 56.4\% & 0.0556 & 0.0285 & 10474.7 & 1.00 \\
\bottomrule
\end{tabular}%
}
\end{table}

Across the three documented IMS failures, the stage-aware curve reduces RMSE by 3.6--18.2\% and MAE by 3.1--31.1\% relative to the best anchored linear approximation. The mean RMSE and MAE reductions are 10.2\% and 15.0\%, respectively; equivalently, the mean RMSE decreases from 0.1141 to 0.1017 and the mean MAE from 0.0909 to 0.0758. Conservative $\Delta$BIC values range from 128.8 to 368.1, favouring the stage-aware representation even after assigning five effective parameters. Because BIC relies on an independent-error likelihood while adjacent snapshots are serially dependent, these values are treated as descriptive model-selection evidence and are interpreted together with the blocked bootstrap.

The bootstrap evidence is more cautious. The 95\% intervals for the mean squared-error gain are $[-0.00207,0.00252]$, $[-0.00126,0.00268]$, and $[-0.00364,0.01120]$ for IMS1-B3, IMS1-B4, and IMS2-B1, respectively. All cross zero. Thus, the official runs consistently favour the stage-aware curve in point estimates and BIC, but the present 300-repetition blocked resampling does not establish uniformly significant improvement. The probability of a positive gain is 0.49, 0.81, and 0.83, showing that support is trajectory dependent.

\begin{figure*}[!t]
    \centering
    \includegraphics[width=\textwidth]{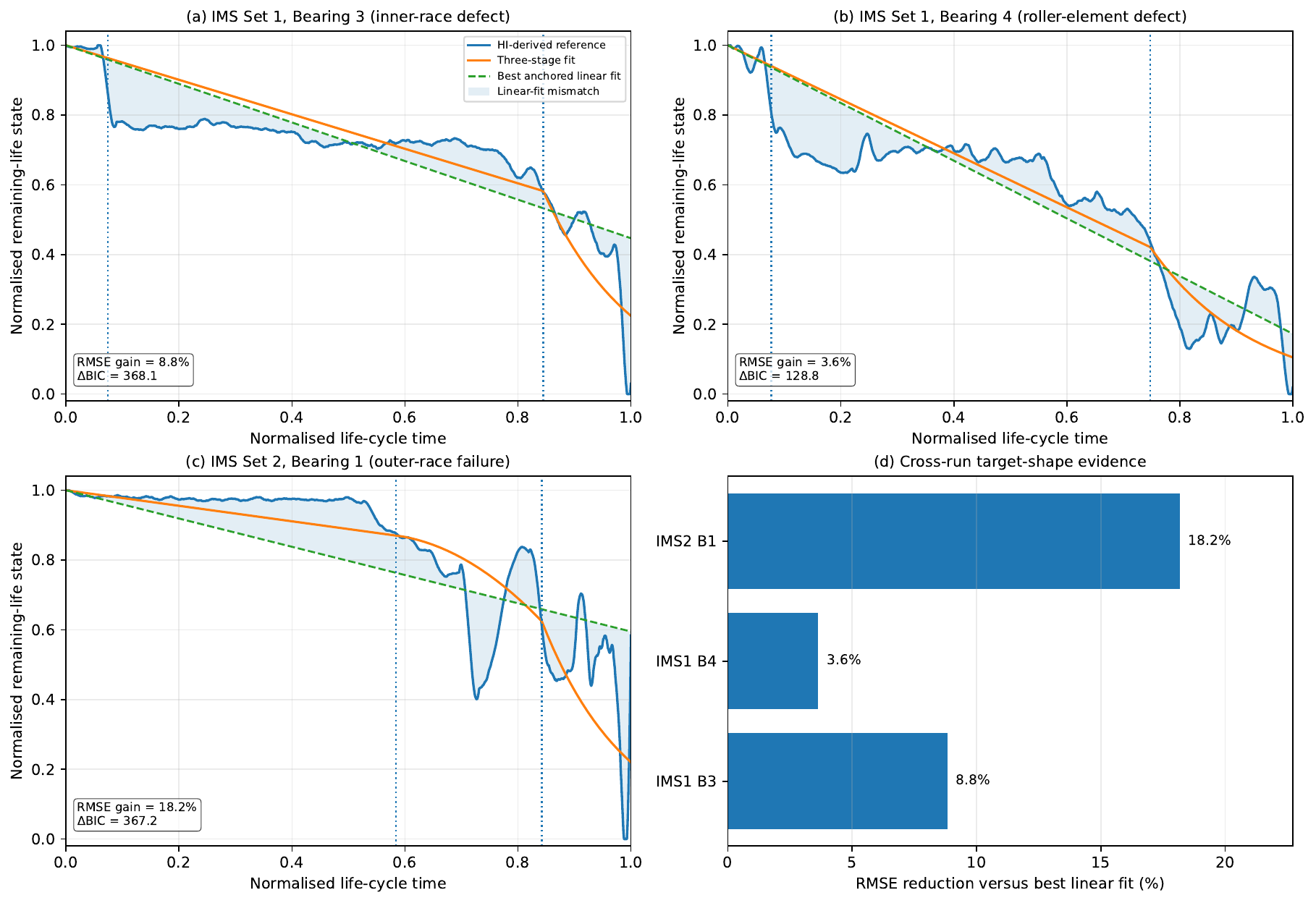}
    \caption{External IMS target-shape assessment on the three documented failed-bearing trajectories. Panels (a)--(c) compare the vibration-HI reference with the continuous three-stage curve and the best anchored linear fit; dotted vertical lines indicate repaired stage boundaries. Panel (d) summarises RMSE reduction. The comparison uses the best fitted linear baseline, making the test stricter than comparison with $1-u$ alone.}
    \label{fig:ims_official}
\end{figure*}

Stagewise errors show why a single aggregate number is insufficient. IMS1-B3 obtains its largest improvement in the late stage (21.1\% RMSE reduction). IMS2-B1 improves strongly in the early and late stages (56.6\% and 16.7\%) but is worse in the middle stage ($-14.5\%$), demonstrating that the chosen functional form is not uniformly superior at every interval. IMS1-B4 shows smaller gains and a slight early-stage deterioration. The evidence therefore supports stage dependence without implying that the specific linear--quadratic--exponential parameterisation is optimal for every bearing.

\begin{figure}[htbp]
    \centering
    \includegraphics[width=1\linewidth]{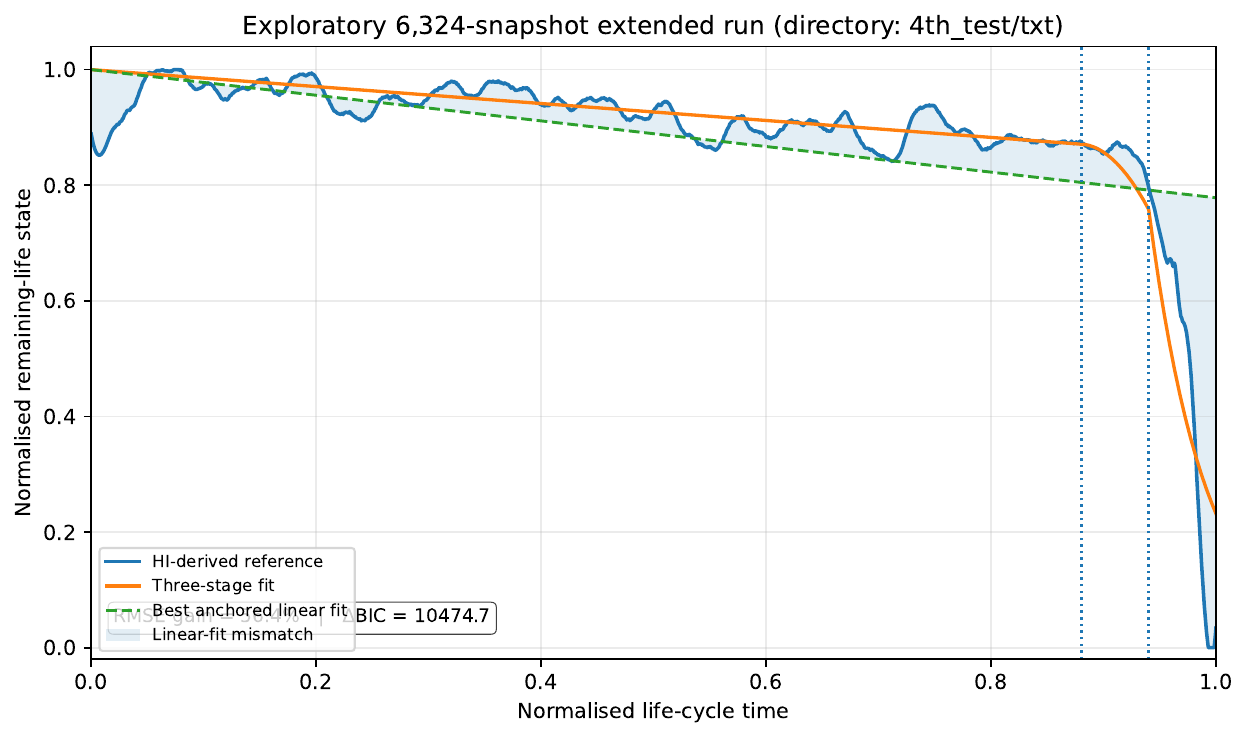}
    \caption{Exploratory analysis of the 6,324-snapshot extended directory. The result strongly favours a stage-aware target, but its provenance does not match the supplied IMS Set~3 documentation and is therefore not treated as confirmatory evidence.}
    \label{fig:ims_extended}
\end{figure}

The exploratory extended run yields a 56.4\% RMSE reduction, 48.8\% MAE reduction, $\Delta$BIC of 10474.7, and a bootstrap interval $[0.00053,0.01474]$ entirely above zero. Figure~\ref{fig:ims_extended} shows the long quasi-stable period and sharp terminal transition. Because the directory does not match the supplied IMS documentation, these values are not used in the confirmatory range reported in the abstract.

\subsection{RQ2: XJTU-SY Held-Out-Bearing Prediction}
Each XJTU-SY one-minute acquisition contains 32,768 samples from horizontal and vertical accelerometers at 25.6~kHz, corresponding to 1.28~s of vibration \cite{xjtuDataset,xjtuTutorial2019}. The three operating conditions are listed in Table~\ref{tab:xjtu_detail}. The predicted quantity is the normalised stage-aware surrogate defined in Eq.~\eqref{eq:piecewiseRUL}; the errors are therefore dimensionless and should not be read directly as minutes or hours.

\begin{table}[h!]
    \centering
    \caption{XJTU-SY operating conditions used in the predictive benchmark}
    \label{tab:xjtu_detail}
    \resizebox{\columnwidth}{!}{%
    \begin{tabular}{cccc}
        \toprule
        \textbf{Condition} & \textbf{Load (kN)} & \textbf{Speed (r min$^{-1}$)} & \textbf{Bearings} \\
        \midrule
        1 & 12 & 2100 & Bearing1-1--Bearing1-5 \\
        2 & 11 & 2250 & Bearing2-1--Bearing2-5 \\
        3 & 10 & 2400 & Bearing3-1--Bearing3-5 \\
        \bottomrule
    \end{tabular}%
    }
\end{table}

For each operating condition, Bearings~1--4 are assigned to development and Bearing~5 is reserved for final evaluation (Table~\ref{tab:dataset_config}). Thus, \texttt{Bearing1\_5}, \texttt{Bearing2\_5}, and \texttt{Bearing3\_5} do not contribute to fitted normalisation statistics, feature-screening thresholds, network parameters, early stopping, hyperparameter choices, or OWA weights.

\begin{table}[htbp]
    \centering
    \caption{XJTU-SY development and final test bearings}
    \label{tab:dataset_config}
    \resizebox{\columnwidth}{!}{%
    \begin{tabular}{ll}
        \toprule
        \textbf{Development bearings} & \textbf{Final test bearing} \\
        \midrule
        Bearing1\_1--Bearing1\_4 & Bearing1\_5 \\
        Bearing2\_1--Bearing2\_4 & Bearing2\_5 \\
        Bearing3\_1--Bearing3\_4 & Bearing3\_5 \\
        \bottomrule
    \end{tabular}%
    }
\end{table}

Table~\ref{test_metrics} reports only the held-out-bearing results because random within-bearing development splits are not independent evidence. The Transformer is the stronger individual branch. OWA reduces MSE from 0.0038 to 0.0037 and increases $R^2$ from 0.9606 to 0.9617, while MAE remains 0.0392 at four-decimal precision. The fusion gain is therefore incremental under this split and should not be interpreted as universal ensemble superiority.

\begin{table}[htbp]
    \centering
    \caption{Held-out XJTU-SY performance on the normalised stage-aware target}
    \label{test_metrics}
    \begin{tabular}{lcccc}
        \toprule
        \textbf{Model} & \textbf{MSE} & \textbf{RMSE} & \textbf{MAE} & \textbf{$R^2$} \\
        \midrule
        CNN--LSTM & 0.0057 & 0.0755 & 0.0423 & 0.9409 \\
        Transformer & 0.0038 & 0.0616 & \textbf{0.0392} & 0.9606 \\
        OWA fusion & \textbf{0.0037} & \textbf{0.0608} & \textbf{0.0392} & \textbf{0.9617} \\
        \bottomrule
    \end{tabular}
\end{table}

\begin{table}[htbp]
\centering
\caption{Component-level ablation on the held-out XJTU-SY bearings}
\label{tab:component_ablation}
\resizebox{\columnwidth}{!}{%
\begin{tabular}{lccc}
\toprule
\textbf{Configuration} & \textbf{MSE} & \textbf{MAE} & \textbf{$R^2$} \\
\midrule
CNN--LSTM only & 0.0057 & 0.0423 & 0.9409 \\
Transformer only & 0.0038 & 0.0392 & 0.9606 \\
CNN--LSTM + Transformer + OWA & \textbf{0.0037} & \textbf{0.0392} & \textbf{0.9617} \\
\bottomrule
\end{tabular}%
}
\end{table}

\begin{figure}[htbp]
    \centering
    \includegraphics[width=1\linewidth]{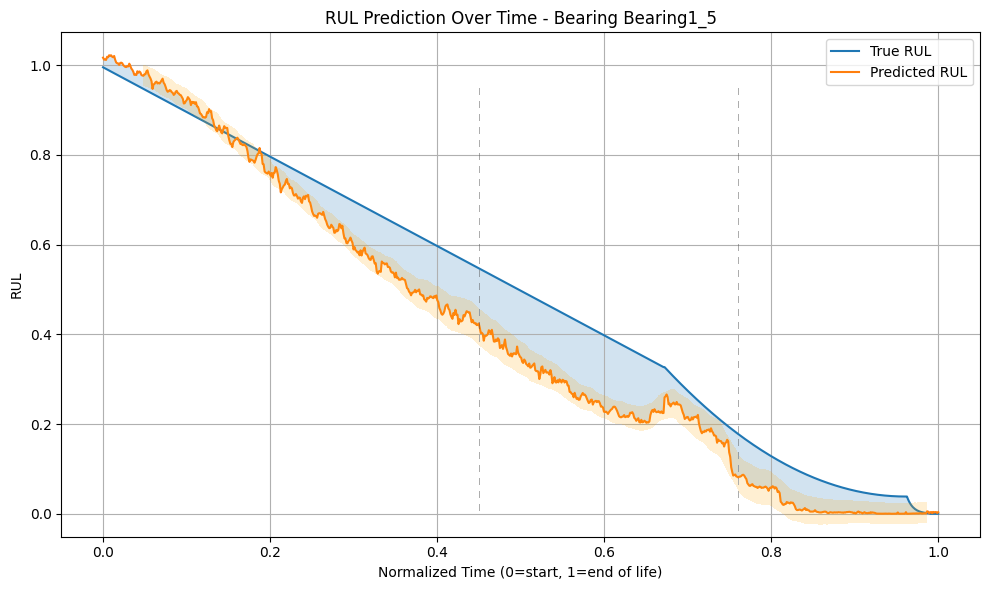}
    \caption{RUL prediction for Bearing1\_5 with empirical residual envelope and stage-boundary markers.}
    \label{fig:5}
\end{figure}

\begin{figure}[htbp]
    \centering
    \includegraphics[width=1\linewidth]{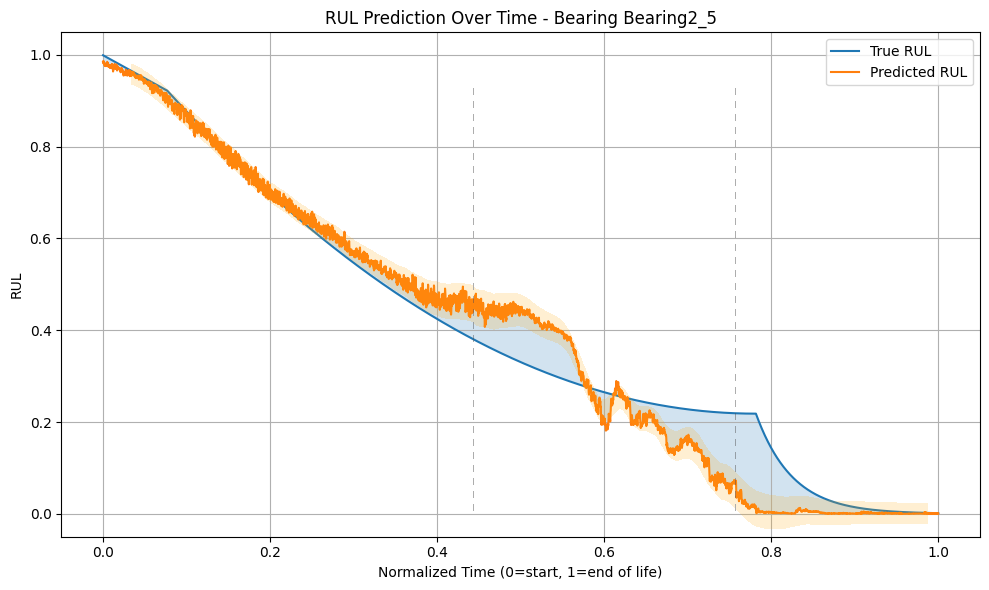}
    \caption{RUL prediction for Bearing2\_5 with empirical residual envelope and stage-boundary markers.}
    \label{fig:6}
\end{figure}

\begin{figure}[htbp]
    \centering
    \includegraphics[width=1\linewidth]{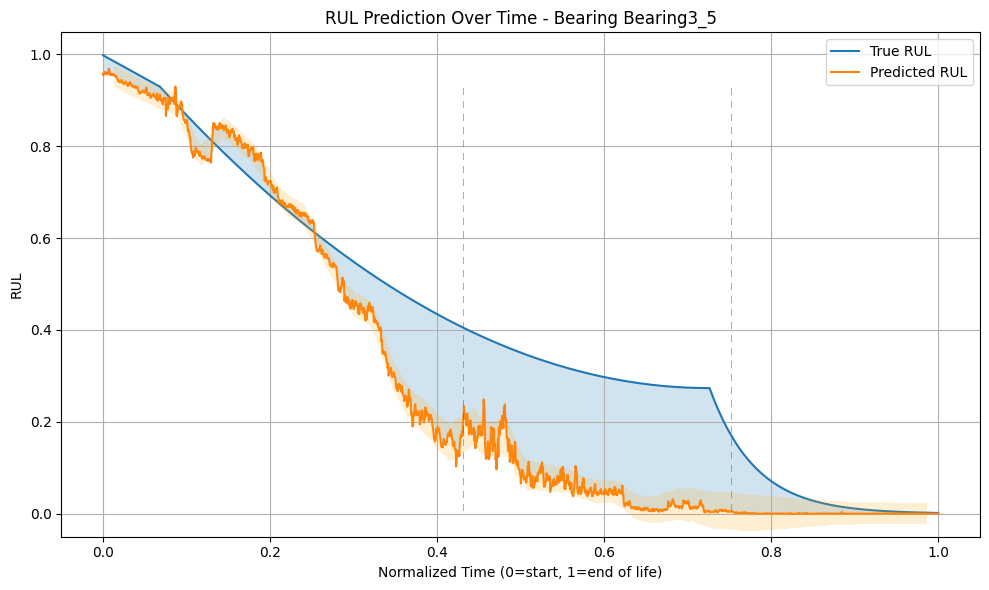}
    \caption{RUL prediction for Bearing3\_5 with empirical residual envelope and stage-boundary markers.}
    \label{fig:7}
\end{figure}

The held-out-bearing test is the relevant generalisation estimate. Random development-window validation is used only for optimisation and OWA fitting because temporally adjacent windows from the same bearing are correlated. Predictions for the three test bearings are shown in Figs.~\ref{fig:5}--\ref{fig:7}; the shaded regions are descriptive residual envelopes and are not calibrated prediction intervals.


\subsection{RQ3: Added Value of Fusion and Limits Under Distribution Shift}
\subsubsection*{Exploratory cross-run target-training ablation}
To test whether target choice affects prediction rather than curve fitting alone, a leave-one-run-out experiment trained the same CNN--LSTM/Transformer/OWA pipeline with either the stage-aware target or the clock-linear target. This experiment used one seed, 25 maximum epochs, patience 5, batch size 64, sequence length 20, and sequence stride 3; it is therefore exploratory rather than a final benchmark. On the three documented IMS runs, stage-aware supervision reduced mean OWA RMSE against the common HI-derived reference from 0.3776 to 0.3372 (10.7\%) and mean MAE from 0.3505 to 0.3028 (13.6\%). However, mean $R^2$ remained negative for both targets, and validation-fitted OWA weights often saturated at or near one branch. The result suggests that a more suitable target helps, but does not remove substantial cross-run and cross-fault distribution shift.

\begin{table}[htbp]
\centering
\caption{Exploratory one-seed IMS cross-run target ablation using OWA fusion; averages are over the three documented failed-bearing test runs.}
\label{tab:ims_target_ablation}
\resizebox{\columnwidth}{!}{%
\begin{tabular}{lccc}
\toprule
\textbf{Training target} & \textbf{Mean RMSE vs. reference} & \textbf{Mean MAE vs. reference} & \textbf{Mean $R^2$} \\
\midrule
Clock-linear $1-u$ & 0.3776 & 0.3505 & $-3.8550$ \\
Stage-aware & \textbf{0.3372} & \textbf{0.3028} & $-2.2988$ \\
\bottomrule
\end{tabular}%
}
\end{table}

\section{Discussion}
\label{sec:discussion}
\subsection{Answers to the Research Questions}
\label{sec:agnostic_cmp}
\textbf{RQ1:} On all three documented IMS failures, the stage-aware curve has lower RMSE and MAE than the best anchored global line, with the largest gains generally appearing where degradation changes fastest. This supports stage dependence of the vibration-derived reference. The bootstrap intervals crossing zero and the small number of complete runs prevent a stronger inferential claim.

\textbf{RQ2:} The XJTU-SY result shows that the retrospective target is learnable from causal feature sequences on the predefined held-out bearings. The Transformer supplies most of the accuracy. The prediction result validates learnability of the constructed target; it does not validate absolute time-to-failure calibration.

\textbf{RQ3:} OWA adds only a small correction to the stronger Transformer branch. The exploratory IMS ablation suggests that stage-aware supervision can reduce mean error, but negative mean $R^2$ and branch-saturated OWA weights show that target choice does not solve cross-run and cross-fault shift.

\subsection{Protocol-Aware Positioning}
Cross-paper RUL comparisons are easily misleading because the reported number depends on the target definition, units, failure threshold, censoring rule, evaluation horizon, and bearing split. Table~\ref{tab:protocol_comparison} therefore reports representative recent results together with their protocols. It is a contextual comparison, not a numerical ranking. The XJTU-SY result here is a dimensionless error against a retrospectively constructed surrogate; RULSurv predicts minutes under censoring; the published IMS Weibull study predicts normalised life percentage on a specified held-out run; and the IMS target-shape rows measure within-run approximation rather than online prediction.

\begin{table}[!t]
\centering
\scriptsize
\caption{Protocol-aware comparison with representative recent XJTU-SY and IMS RUL studies. Metrics are not directly rankable across rows because targets, units, splits, and task definitions differ.}
\label{tab:protocol_comparison}
\begin{tabularx}{\textwidth}{@{}p{0.09\textwidth}p{0.16\textwidth}p{0.25\textwidth}p{0.18\textwidth}X@{}}
\toprule
\textbf{Dataset} & \textbf{Study/method} & \textbf{Evaluation protocol} & \textbf{Target/output} & \textbf{Reported result and interpretation} \\
\midrule
XJTU-SY & RULSurv, Random Survival Forest \cite{lillelund2025rulsurv} & Five-fold cross-validation; highest-load condition C1; 25\% random censoring & Time to failure in minutes with a survival distribution & MAE $12.6$~min (95\% CI $11.8$--$13.4$) and mean CRA $0.7586$ over five bearings. Stronger uncertainty/censoring treatment, but not comparable numerically with a normalised surrogate. \\
XJTU-SY & Proposed CNN--LSTM + Transformer + OWA & Bearings~1--4 for development and Bearing~5 for final testing within each of the three conditions & Continuous normalised stage-aware degradation-state surrogate & RMSE $0.0608$, MAE $0.0392$, and $R^2=0.9617$. This is the main held-out predictive result of the present study. \\
IMS & Weibull-informed neural network \cite{vonhahn2022weibull} & Train: Run~2 Bearing~1 and Run~3 Bearing~3; validation: Run~1 Bearing~3; test: Run~1 Bearing~4; horizontal channel & Normalised life-percentage RUL & On the specified test run, RMSE $0.146$ and $R^2=0.735$. This is the closest published IMS predictor comparison located, but it uses a different HI, target, channel selection, and split. \\
IMS & Proposed stage-aware target fit (official runs) & Retrospective fitting on IMS1-B3, IMS1-B4, and IMS2-B1; no cross-run predictor training in this row & Vibration-HI-derived remaining-life reference & Mean stage RMSE $0.1017$ and mean $R^2=0.7196$; mean RMSE gain $10.2\%$ and MAE gain $15.0\%$ over the best anchored linear fit. This row supports target shape, not online generalisation. \\
IMS & Proposed OWA predictor, exploratory & Leave-one-run-out over the three documented failed-bearing runs; one seed and 25 maximum epochs & Common HI-derived reference; stage-aware training target & Mean RMSE $0.3372$, MAE $0.3028$, and $R^2=-2.2988$. The weak absolute transfer result is reported to show that nonlinear supervision does not by itself solve domain shift. \\
\bottomrule
\end{tabularx}
\end{table}

Table~\ref{tab:literature_position} complements the numeric comparison by locating the proposed method among recent stage-aware, hybrid temporal, transfer-learning, survival, and probabilistic approaches. ``Adds'' denotes a distinct auditable capability under the present protocol, not universal superiority.

\begin{table}[!t]
\centering
\scriptsize
\caption{Methodological positioning relative to representative recent work on bearing RUL.}
\label{tab:literature_position}
\begin{tabularx}{\textwidth}{@{}>{\raggedright\arraybackslash}p{0.14\textwidth}>{\raggedright\arraybackslash}p{0.16\textwidth}>{\raggedright\arraybackslash}p{0.22\textwidth}>{\raggedright\arraybackslash}X@{}}
\toprule
\textbf{Reference} & \textbf{Dataset} & \textbf{Main mechanism} & \textbf{Relation to the proposed study} \\
\midrule
Qiu \emph{et al.} \cite{qiu2023piecewise} & Bearing benchmark & Piecewise RUL estimation with a temporal convolutional network & Both recognise stage-dependent degradation. The present method adds two separately auditable experts and validation-constrained decision fusion; published protocols remain non-interchangeable. \\
Peng \emph{et al.} \cite{peng2023let} & XJTU-SY & Local-enhancing Transformer with temporal convolutional attention & Their local enhancement is integrated inside one Transformer pipeline; ours keeps local CNN--LSTM and long-context Transformer outputs separate for branch-level ablation and rank-based fusion. \\
Tang \emph{et al.} \cite{tang2024parallel} & Bearing benchmark & Parallel TCN and Transformer representation learning & Architecturally close to the local/global motivation. The distinguishing elements here are explicit target construction, chronological stage repair, and OWA at decision level. \\
Lillelund \emph{et al.} \cite{lillelund2025rulsurv} & XJTU-SY & Censoring-aware survival analysis and probabilistic RUL & Provides RUL in minutes and explicit censoring support. The present study instead focuses on an interpretable degradation-state surrogate and separable expert fusion. \\
von Hahn and Mechefske \cite{vonhahn2022weibull} & IMS & Weibull knowledge embedded in a neural-network loss & Their work integrates reliability knowledge into training and reports held-out IMS prediction. Ours contributes explicit three-stage target analysis and cross-dataset target-shape evidence. \\
Xu \emph{et al.} \cite{xu2025domain} & XJTU-SY & HI-weighted subdomain alignment for cross-condition transfer & Directly addresses domain shift and is therefore stronger for cross-condition adaptation; such adaptation remains a limitation of the current predictor. \\
Bott \emph{et al.} \cite{bott2026uncertainty} & Simulated and experimental bearings & Physics-based simulation; conditional normalising flows & Provides a principled predictive distribution. The residual envelopes used here are descriptive and are not a substitute for calibrated uncertainty. \\
\bottomrule
\end{tabularx}
\end{table}

The combined evidence supports two bounded conclusions. On XJTU-SY, the Transformer captures most of the predictive strength and the CNN--LSTM supplies a small complementary correction through OWA. On IMS, the most reproducible cross-dataset result concerns target shape: the stage-aware surrogate fits the vibration-derived reference better than the best global linear approximation on all documented failed-bearing runs, although blocked-bootstrap strength varies. The exploratory IMS predictor remains weak under cross-run transfer. Consequently, the contribution is the coordinated and auditable combination of explicit stage-target construction, separable local and long-context experts, validation-fitted rank fusion, and transparent two-dataset validation---not a claim of universal state-of-the-art dominance.
\subsection{Why the Contribution Is Target-Centred Rather Than Architecture-Centred}
The experimental pattern does not support presenting the CNN--LSTM/Transformer combination as the main novelty. The Transformer already achieves nearly all of the held-out XJTU accuracy, and the OWA gain is small. The stronger contribution is methodological: it exposes target construction as a testable modelling choice, compares it with a fitted linear alternative on a second rig, and separates descriptive target adequacy from predictive generalisation. This framing is also consistent with recent health-indicator and physics-informed prognostics literature, where degradation representation, uncertainty, censoring, and transfer are treated as distinct design problems \cite{zhou2022,li2024review,guo2024wiener,lillelund2025rulsurv}.

\subsection{Practical Interpretation}
The proposed output should be read as a normalised degradation-state score. It can support condition tracking or act as an input to a later time-calibration layer, but it does not by itself represent minutes or hours remaining. Deployment would require a machine-specific failure threshold, mapping from state to time under the current operating regime, uncertainty calibration, and drift monitoring. In this sense, the present work addresses one layer of a prognostic system: constructing and learning a condition-consistent target.

\section{Limitations and Deployment Considerations}
\label{sec:limits_deploy}
The first limitation is experimental independence. The XJTU-SY benchmark uses one predefined bearing-wise hold-out and one reported training realisation. This is stronger than random window-level testing, but it does not quantify variation over alternative bearing assignments, seeds, fault modes, or operating conditions. Random development-window validation is retained only for early stopping and fusion fitting and may be optimistic because neighbouring windows are correlated.

Second, the IMS target-shape assessment contains only three documented failed-bearing trajectories. It is external in the sense of using a different rig, but each target is fitted retrospectively within its own complete run. The assessment therefore tests functional adequacy, not prospective generalisation of fixed target parameters. BIC is supportive but inherits assumptions about the residual likelihood; the moving-block bootstrap provides the more conservative dependence-aware check, and its intervals cross zero for all three official runs.

Third, the target depends on full-run health-indicator orientation, smoothing, stage repair, and end-of-run normalisation. These operations are suitable for retrospective supervision but are unavailable for a live, incomplete trajectory unless replaced by online change detection and fixed calibration learned from historical runs. The output is a degradation-state surrogate, not a direct measurement of physical time remaining.

Fourth, the exploratory IMS target-training ablation uses one seed, 25 maximum epochs, and a small set of heterogeneous failed runs. Its negative mean $R^2$ values and frequent OWA saturation are evidence of unresolved domain shift, not merely insufficient fusion. Stronger transfer experiments require repeated grouped evaluation, fault-aware or condition-aware adaptation, and more diverse training runs.

Fifth, the arXiv source package is intended for manuscript compilation and does not include every XJTU implementation artefact needed for exact independent reproduction, including the final selected feature list, raw-signal window/step, smoothing span, dropout and weight-decay values, sequence length, and numerical OWA weights. This is a reproducibility limitation; those configuration objects should be released separately with the executable code rather than reconstructed from incomplete records.

Finally, the residual envelopes in the prediction figures are descriptive. They are not coverage-calibrated prediction intervals. Conformal prediction or an explicit probabilistic model should be evaluated before uncertainty is used for maintenance-risk decisions \cite{angelopoulos2023conformal,bott2026uncertainty}.

After offline fitting, inference is lightweight: apply frozen preprocessing, compute selected trailing-window descriptors, evaluate two compact predictors, and combine two scalar outputs. A real installation must also include sensor-quality checks, missing-channel handling, latency and memory profiling, operating-condition detection, drift alarms, scheduled recalibration, conservative decision thresholds, and validation on the intended machine.

\section{Conclusion and Future Work}
\label{CFW}
This study addressed a frequently overlooked assumption in bearing prognostics: the regression label itself. A clock-linear RUL label is convenient, but it can disagree with vibration-observed degradation that remains weak for a long interval and accelerates near failure. The proposed framework therefore separates target construction from prediction. It builds an oriented health indicator, identifies a chronological three-stage degradation sequence, fits a continuous phenomenological target, and learns that target with causal CNN--LSTM and Transformer sequences combined by validation-fitted OWA.

On the predefined XJTU-SY held-out bearings, the fused predictor achieved RMSE 0.0608, MAE 0.0392, and $R^2=0.9617$. The Transformer was the dominant branch and OWA supplied only a modest incremental benefit. On the three documented IMS failed-bearing trajectories, the stage-aware curve reduced RMSE by 3.6--18.2\% and MAE by 3.1--31.1\% relative to the best anchored linear fit. Conservative BIC differences favoured the stage-aware representation, whereas dependence-aware bootstrap intervals crossed zero. The evidence is therefore consistent and practically meaningful as a descriptive target-shape result, but not sufficient for a universal statistical or physical claim.

The negative exploratory transfer result is equally important: a more condition-consistent target does not by itself overcome cross-run and cross-fault distribution shift. The defensible conclusion is specific. For the evaluated trajectories, stage-dependent degradation-state labels align with the vibration-derived reference better than a single global line, and that target can be predicted on a predefined XJTU bearing hold-out. The study does not establish a universal nonlinear law for clock-time RUL, calibrated uncertainty, or robust cross-domain deployment.

The highest-priority next steps are repeated group-wise XJTU evaluation over multiple seeds and held-out assignments; verification of the extended IMS directory; comparison with monotone splines, logistic/Gompertz curves, and change-point models; online stage detection without full-run information; domain-adaptive prediction; and calibrated predictive intervals. Exact preprocessing objects, feature rankings, stage boundaries, seeds, and fusion weights should be included in a separate reproducibility release.

\FloatBarrier
\section*{Acknowledgements}
	This work was supported by the Vice Chancellor for Research and Technology of Sharif University of Technology under Grant No.~G4032104. Hardware required for simulations in this article was partially provided by AI Ahura.

\section*{Data Availability}
XJTU-SY and IMS are publicly available benchmark datasets through their original distributors. The arXiv source package contains only the manuscript source and figures required for compilation. Processed analysis artefacts, executable code, and the complete XJTU configuration are not included in this source archive and should be released separately with a persistent identifier.

\section*{Declaration of competing interest}
The authors declare that they have no known competing financial interests or personal relationships that could have appeared to influence the work reported in this paper.

\section*{Declaration of generative AI and AI-assisted technologies in the manuscript preparation process}
During the preparation of this work, the authors used OpenAI ChatGPT to assist with language editing, journal-template conversion, and organisation of the related-work discussion. The authors reviewed and edited all generated material, independently verified the cited sources and technical statements, and take full responsibility for the content of the article.

\FloatBarrier

\end{document}